\documentclass[journal]{IEEEtran}
\ifCLASSINFOpdf
\else
\fi

\usepackage[hyphens]{url}
\usepackage[hidelinks]{hyperref}
\hypersetup{breaklinks=true}
\usepackage{cite}
\usepackage{amsmath,amssymb,amsfonts}
\usepackage{graphicx}
\usepackage{textcomp}
\usepackage{xcolor}
\usepackage{multicol}
\usepackage{multirow}
\usepackage{afterpage}
\usepackage{siunitx}
\usepackage{stfloats}
\usepackage{cuted}

\def\BibTeX{{\rm B\kern-.05em{\sc i\kern-.025em b}\kern-.08em
    T\kern-.1667em\lower.7ex\hbox{E}\kern-.125emX}}

\usepackage[ruled]{algorithm2e}
\usepackage{algpseudocode}

\usepackage{array}
\usepackage{textcomp}
\usepackage{makecell}
\usepackage{hhline}
\usepackage[flushleft]{threeparttable}

\usepackage{multicol}
\usepackage{multirow}
\usepackage{booktabs}

\usepackage{verbatim}

\usepackage{url}

\newcolumntype{M}[1]{>{\centering\arraybackslash}m{#1}}
\newcolumntype{P}[1]{>{\centering\arraybackslash}p{#1}}

\newcommand{\trp}[1]{{#1}^{\mathsf{T}}}

\newcommand{\thickhline}{
    \noalign {\ifnum 0=`}\fi \hrule height 1pt
    \futurelet \reserved@a \@xhline
}
\DeclareMathOperator*{\argmax}{arg\,max}
\DeclareMathOperator*{\argmin}{arg\,min}

\makeatletter
\def\env@dmatrix{\hskip -\arraycolsep
  \let\@ifnextchar\new@ifnextchar
  \extrarowheight=2ex
  \array{*\c@MaxMatrixCols{>{\displaystyle}c}}}

\makeatother

\newtheorem{definition}{Definition}
\newtheorem{theorem}{Theorem}

\usepackage{blkarray}

\begin{document}

\title{Rigidity-Based Multi-UAV Trajectory Optimization for Rapid Cooperative Emergency Target Localization}

\author{Halim~Lee and Jiwon~Seo,~\IEEEmembership{Senior Member,~IEEE}
\thanks{Copyright (c) 2026 IEEE. Personal use of this material is permitted. However, permission to use this material for any other purposes must be obtained from the IEEE by sending a request to pubs-permissions@ieee.org.}
\thanks{Manuscript received March 00, 2026; }
\thanks{This work was supported in part by the National Research Foundation of Korea (NRF), funded by the Ministry of Science and ICT (MSIT), under Grants RS-2024-00358298 and RS-2025-25393156; 
in part by the Korea Aerospace Administration (KASA), under Grant RS-2022-NR067078;  
in part by Grant RS-2024-00407003 from the ``Development of Advanced Technology for Terrestrial Radionavigation System'' project, funded by the Ministry of Oceans and Fisheries, Republic of Korea; 
in part by the Unmanned Vehicles Core Technology Research and Development Program through the NRF and the Unmanned Vehicle Advanced Research Center (UVARC), funded by MSIT, Republic of Korea, under Grant RS-2020-NR046546; 
and in part by the Institute of Information \& Communications Technology Planning \& Evaluation (IITP) through the Information Technology Research Center (ITRC) program, funded by MSIT, under Grant IITP-2025-RS-2024-00437494.
\textit{(Corresponding Author: Jiwon Seo.)}}
\thanks{Halim~Lee and Jiwon~Seo are with the School of Integrated Technology, Yonsei University, Incheon 21983, Republic of Korea (e-mail: halim.lee@yonsei.ac.kr; jiwon.seo@yonsei.ac.kr).}
}

%
%

\markboth{}%
{Lee \MakeLowercase{\textit{and}} Seo: Rigidity-Based Multi-UAV Trajectory Optimization for Rapid Cooperative Emergency Target Localization}
%



\maketitle

\begin{abstract} 
Reducing the response time required for accurate localization of emergency callers is a critical challenge in vehicular technology and public-safety networks, where timely and reliable positioning directly affects mission outcomes.
Although mobile devices commonly rely on global navigation satellite systems (GNSS), Wi-Fi, or cellular positioning, their accuracy and availability can be degraded by signal reception conditions, infrastructure coverage, and regulatory constraints. 
Unmanned aerial vehicle (UAV)-based localization has therefore emerged as a promising alternative, where UAVs act as airborne sensors to cooperatively estimate the target position. 
However, existing Fisher information matrix (FIM)-based UAV trajectory optimization methods rely on the current target estimate and can be less effective in the early mission stage, when measurement diversity is limited and the estimate is highly uncertain. 
To address this problem, we propose a rigidity-based UAV trajectory optimization method that maximizes the smallest nonzero singular value of the rigidity matrix associated with the UAV--target sensing graph, thereby improving geometric conditioning and reducing position ambiguity. 
We further introduce a pruning-based rigidity matrix reduction strategy for efficient real-time implementation. 
Simulation results show that the proposed method reduces search time by 32.9\% compared with existing FIM-based methods while satisfying the Federal Communications Commission (FCC) horizontal emergency localization requirement within a shorter timeframe. 
Additional evaluations confirm its scalability, robustness to UAV positioning errors and non-line-of-sight (NLOS) path loss, low sensitivity to heading-angle parameters, computational feasibility, low communication overhead, and more graceful performance degradation than proximal policy optimization (PPO)-based learning baselines under severe sensing and navigation perturbations.
\end{abstract}


%
\IEEEpeerreviewmaketitle

\section{Introduction}
\label{sec:Introduction}

\IEEEPARstart{T}{he} rapid and accurate localization of a target in emergency situations is a critical requirement in vehicular and mission-critical wireless systems, where reliable positioning directly affects safety-of-life services. 
In 2018, the Federal Communications Commission (FCC) of the United States estimated that ``as many as 10,000 lives could be saved each year if the 911 emergency dispatching system were able to reach callers just one minute faster'' \cite{FCC18}. 
Given that even small delays in emergency response can have severe consequences, minimizing the time required to obtain reliable location information is essential.

Currently, the FCC mandates that mobile radio service providers ensure horizontal positioning accuracy within $\pm$50~m for 80\% of all wireless 911 calls \cite{FCC20}. 
In many regions, most mobile devices rely on global navigation satellite systems (GNSS) to report their locations during emergency calls. 
However, GNSS performance can be significantly degraded in urban environments due to signal blockage and multipath reflections caused by buildings \cite{Lee22:Urban, Lee23:Nonlinear}. 
In addition, GNSS signals are vulnerable to radio-frequency interference because of their weak received signal power \cite{Kim22:First, Park21:Single, Son19:Universal}, and ionospheric anomalies can further degrade GNSS accuracy \cite{Lee17:Monitoring, Lee22:Optimal, Sun21:Markov}. 
Furthermore, in regions where emergency positioning services are not mandated or fully deployed, first responders may not have access to the location information of certain mobile devices.

Cellular or Wi-Fi-based positioning methods \cite{Cheng05:Accuracy, Win11:Network, Lee20:Preliminary, Moon24:HELPS} can serve as alternatives to GNSS. 
However, these approaches often suffer from limited geometric diversity because they rely on fixed infrastructure such as cellular base stations or access points, which can restrict localization accuracy. 
To address these limitations, unmanned aerial vehicles (UAVs) equipped with Internet-of-Things (IoT) devices have emerged as a promising platform for target localization \cite{Lee22:Evaluation, Lee23:Performance, Wang18:Performance, Uluskan20:Noncausal, Koohifar16:Receding, Xia24:Eye}. 
Unlike stationary infrastructure, UAVs can maneuver to collect measurements from geometrically diverse viewpoints, thereby improving localization performance.

UAV-based localization systems can utilize various types of signal measurements, including time-of-arrival (TOA), angle-of-arrival (AOA), Doppler shift, and received signal strength (RSS) \cite{Dogancay12:UAV, Nguyen16:Optimal, Lee13:Cooperative, Wang18:Performance, Uluskan20:Noncausal, Koohifar16:Receding}. 
Among these options, RSS measurements provide a particularly attractive solution because they require low hardware complexity and can be readily obtained using lightweight radio modules on resource-constrained UAV platforms \cite{Zanella16:Best}. 
For this reason, RSS-based localization has been widely investigated in UAV-assisted sensing systems.

The performance of cooperative localization is strongly influenced by the trajectories of the UAVs collecting measurements. 
Previous studies have therefore investigated UAV trajectory optimization using the Fisher information matrix (FIM) as a key design metric \cite{LeCadre97:Discrete, Passerieux98:Optimal, He19:Trajectory, He19:The, Dogancay12:UAV, Nguyen16:Optimal, Lee13:Cooperative, Wang18:Performance, Uluskan20:Noncausal, Koohifar16:Receding}. 
The FIM provides a measure of the expected information content of the measurements and is commonly used to derive the Cram\'er--Rao lower bound (CRLB), which represents a theoretical lower bound on the estimation error covariance. 
Consequently, several studies have proposed trajectory optimization methods that maximize the determinant of the FIM to minimize the uncertainty of the target position estimate.

However, FIM-based trajectory optimization implicitly relies on the availability of a reasonably accurate operating point. 
The FIM is evaluated using the current estimate of the target position, which is inferred from noisy measurements. 
In the early stages of an emergency localization mission, only a small number of measurements may be available, and UAVs may initially start from a common base or vertiport. 
Under these conditions, the inferred target estimate may deviate significantly from the true location, causing the FIM to reflect only the local curvature of the likelihood around an off-truth estimate. 
As a result, FIM-guided trajectory updates may become less effective during the critical early phase of the mission when geometric diversity is still limited.

In such situations, the localization problem often exhibits \emph{position ambiguity}. 
Such ambiguity arises when multiple candidate target locations yield similar likelihood values under noisy RSS measurements, making it difficult to identify the true target location uniquely. 
While collecting additional measurements and increasing geometric diversity can eventually resolve this ambiguity, time-constrained emergency scenarios require methods that can reduce ambiguity as quickly as possible.

To address this challenge, this paper leverages the concept of \emph{rigidity} from graph theory \cite{Sitharam18:Handbook} to design UAV trajectories. Rigidity characterizes whether a geometric framework preserves its structure under distance-preserving deformations. In the considered RSS-based localization problem, the UAVs and the target can be viewed as vertices of a UAV--target sensing framework, while RSS-derived distance-related constraints can be viewed as edges. From this perspective, position ambiguity is closely related to poor geometric conditioning of the sensing framework: when UAV measurements are collected from geometrically similar or near-degenerate viewpoints, multiple target hypotheses may remain difficult to distinguish. Therefore, improving the rigidity of the UAV--target framework provides a natural way to encourage measurement geometries that reduce ambiguity in RSS-based localization.

\begin{figure}
    \centering
    \includegraphics[width=0.7\linewidth]{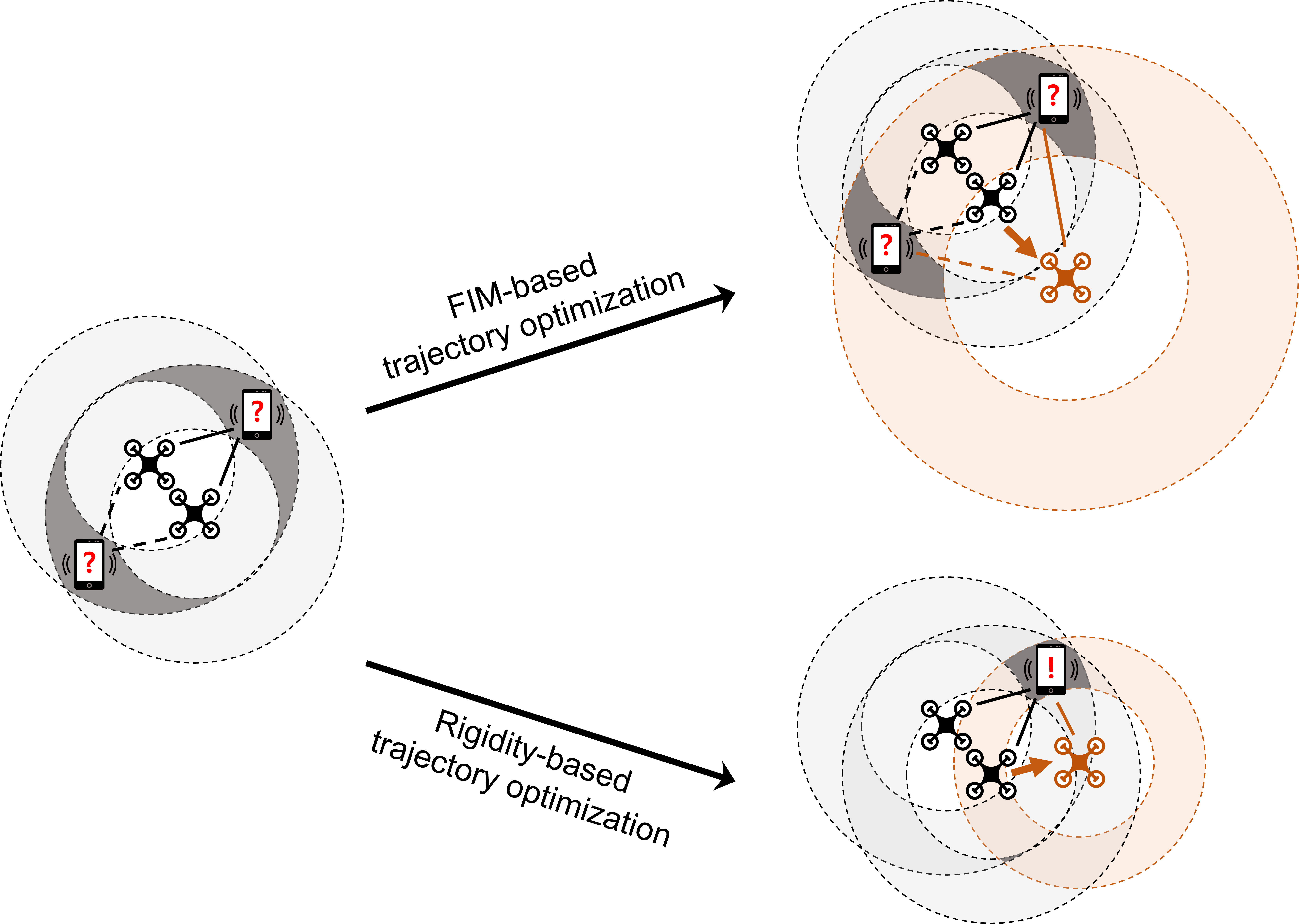}
    \caption{Conceptual illustration of target localization scenarios. The proposed rigidity-based UAV trajectory optimization method (bottom-right) encourages UAVs to move into geometries that more effectively reduce position ambiguity, whereas a conventional FIM-based update (top-right) yields a more limited reduction in this illustrative example. The circular rings around the two UAVs represent distance constraints estimated from RSS measurements, including their associated uncertainties.}
    \label{fig:PositionAmbiguity}
\end{figure}

Fig.~\ref{fig:PositionAmbiguity} conceptually illustrates the intuition behind the proposed rigidity-based trajectory optimization compared with a conventional FIM-based approach. The left-hand figure shows a scenario with significant position ambiguity, where the true target position remains uncertain within overlapping candidate regions derived from RSS measurements. When an additional measurement is collected, the trajectory selected by a FIM-based update may only marginally reduce the ambiguous region, whereas the rigidity-based trajectory encourages a configuration that more effectively reduces the ambiguity.

The unique benefit of the proposed rigidity-based optimization is that it guides UAVs toward non-degenerate sensing geometries that improve structural distinguishability among candidate target locations, especially when the early-stage likelihood surface is ambiguous. This benefit arises from the way the proposed method uses the current target estimate. Existing FIM-based methods use the current MLE as a statistical operating point and evaluate the expected information content of future measurements around that point. Therefore, when the early-stage estimate is biased or the likelihood surface is multimodal, such methods may remain locally informative around an incorrect hypothesis while leaving global position ambiguity unresolved. In contrast, the proposed rigidity-based method uses the current estimate as a target vertex in the UAV--target sensing framework and evaluates the geometric conditioning of this framework through relative displacement constraints. Maximizing the smallest nonzero singular value of the rigidity matrix therefore encourages sensing geometries that are less degenerate and more informative for distinguishing among candidate target locations. To the best of our knowledge, this is the first study to formulate UAV trajectory optimization as the maximization of the smallest nonzero singular value of the rigidity matrix for cooperative localization.
The main contributions of this paper are summarized as follows:

\begin{itemize}

\item We propose a rigidity-based trajectory optimization framework for cooperative UAV target localization. The proposed method formulates the trajectory planning problem by maximizing the smallest nonzero singular value of the rigidity matrix associated with the UAV–target sensing graph, which directly improves the geometric conditioning of the localization problem and rapidly reduces position ambiguity.

\item We develop a rigidity matrix reduction strategy that eliminates less informative vertices using a priority-based pruning mechanism, significantly reducing the computational cost while preserving the dominant geometric constraints required for localization.

\item We conduct comprehensive simulation studies to evaluate the proposed rigidity-based trajectory optimization framework, including comparisons with FIM-based methods and a proximal policy optimization (PPO)-based learning baseline, scalability with respect to the number of UAVs, robustness to UAV positioning errors, sensitivity to additional non-line-of-sight (NLOS) path loss, sensitivity to heading-angle parameters, computational feasibility, and communication overhead.

\end{itemize}

\section{Related Work}

\subsection{UAV Trajectory Optimization for Target Localization}

The trajectories of UAVs play a critical role in the performance of UAV-based target localization. 
Over the past few decades, researchers have explored various optimization metrics for trajectory design (i.e., path planning \cite{Huang20:Multiobjective, Yin17:Offline, Yu22:Novel, Dogancay12:UAV, He19:The}) to enhance localization accuracy and efficiency.

The FIM has long been utilized as an optimization metric \cite{LeCadre97:Discrete, Passerieux98:Optimal, He19:Trajectory, He19:The, Dogancay12:UAV, Nguyen16:Optimal, Lee13:Cooperative, Wang18:Performance, Uluskan20:Noncausal, Koohifar16:Receding, Lee23:Performance}. 
In bearing-only localization, maximizing the determinant of the FIM has been considered \cite{LeCadre97:Discrete, Passerieux98:Optimal, He19:Trajectory, He19:The}. 
In \cite{Dogancay12:UAV}, the FIM for time difference of arrival (TDOA), angle of arrival (AOA), and scan-based hybrid localization was introduced, and an FIM-based trajectory optimization problem was formulated. 
In \cite{Nguyen16:Optimal}, the optimal geometry for multistatic TOA localization was analyzed. 
Their analysis was also based on the FIM, and UAV paths were optimized to maximize the determinant of the FIM. 
In \cite{Lee13:Cooperative}, cooperative target localization using two small UAVs equipped with heterogeneous sensors (distance-only and bearing-only) was considered. 
The trajectories of the two UAVs were optimized to maximize the determinant of the accumulated FIM. 
In \cite{Wang18:Performance, Uluskan20:Noncausal, Koohifar16:Receding, Lee23:Performance}, FIM-based optimization problems were designed for target localization scenarios in which UAVs collect RSS measurements. 
Wang et al. \cite{Wang18:Performance} derived the FIM under distance-dependent RSS noise. 
Uluskan \cite{Uluskan20:Noncausal} and Koohifar et al. \cite{Koohifar16:Receding} applied noncausal and receding horizon control strategies, respectively, to design FIM-based optimization problems.

As discussed in Section \ref{sec:Introduction}, the FIM is typically evaluated at inferred parameter values because the true state is unknown.
Prior studies \cite{Dogancay12:UAV, Wang18:Performance, Uluskan20:Noncausal} commonly used maximum likelihood estimation (MLE) to obtain the operating point.
However, real-world constraints on localization time and UAV geometric diversity often yield off-truth estimates and multimodal likelihoods.
In such conditions, the FIM evaluated at the inferred state captures only the local curvature of the log-likelihood around that point, which may not fully represent the information associated with the true state.
As a result, FIM-guided trajectory updates can be less effective under high initial ambiguity.

Previous FIM-based UAV localization studies \cite{Wang18:Performance, Uluskan20:Noncausal, Koohifar16:Receding} primarily considered scenarios where the likelihood surface is relatively well-behaved---either due to minimal measurement noise or sufficient geometric diversity---thereby ensuring that the MLE remains reasonably close to the true target location.
Table \ref{tab:RSS_error} provides a comparison between the RSS error assumptions made in previous studies \cite{Wang18:Performance, Uluskan20:Noncausal, Koohifar16:Receding} and those recommended by ITU-R for line-of-sight (LOS) scenarios \cite{ITUR09}. 
Previous studies \cite{Wang18:Performance, Uluskan20:Noncausal, Koohifar16:Receding} have predominantly explored FIM-based UAV trajectory optimization under optimistic RSS error conditions, often assuming minimal errors (e.g., 0.01 dB). 
However, ITU-R recommends considerably higher shadowing variations---typically 4–6~dB in suburban macro or rural macro environments---which leads to a substantially more ambiguous likelihood surface during early mission stages.

\begin{table*}
\footnotesize
\centering
\caption{Comparison of RSS Error Assumptions in the Literature (Reproduced from Table 1 in \cite{Lee23:Performance})} \label{tab:RSS_error}
\vspace{-4mm}
\begin{center}{
\renewcommand{\arraystretch}{1.4}
 \begin{tabular}{ c |P{1.3cm} P{1.3cm} P{1.75cm} P{1.3cm} | P{1.5cm} P{1.3cm} P{1.3cm}}
 \Xhline{1.0pt}
 \rule{0pt}{15pt} {} & \multicolumn{4}{c|}{\thead{ITU-R LOS Scenarios \cite{ITUR09}}} & \multicolumn{3}{c}{\thead{Previous Studies}} \\ 
 \multirow{-2}{*}{\thead{Reference}} & \thead{Urban \\ Micro} & \thead{Urban \\ Macro} & \thead{Suburban \\ Macro} & \thead{Rural \\ Macro} & \thead{\cite{Wang18:Performance}} & \thead{\cite{Uluskan20:Noncausal}} & \thead{\cite{Koohifar16:Receding}} \\
 \hline
 \rule{0pt}{20pt} \thead{RSS Error \\ (dB)} & 3 & 4 & 4--6 & 4--6 & $0.01 \times d$ & \makecell{0.01 \\ or 5} & 0.01\\
 \Xhline{1.0pt}
 \multicolumn{8}{l}{\footnotesize \textit{Note:} The variable $d$ is the distance between the UAV and target.}
\end{tabular}}
\end{center}
\end{table*}

In \cite{Uluskan20:Noncausal}, a 5 dB RSS error scenario was examined under a favorable geometric configuration in which eight UAVs were positioned to surround the target.
However, such well-distributed geometries are not always achievable in practice.
In many real-world missions, UAVs may need to begin their operations from a single base or operate within other deployment constraints, resulting in less diverse geometric configurations.

Our study focuses on scenarios characterized by large RSS errors and unfavorable UAV geometries, which result in more ambiguous target position estimates. 
The principal contribution of this study is the development of a rigidity-based optimization approach that addresses the challenge of reducing position ambiguity in realistic scenarios by accounting for both practical RSS errors and realistic UAV geometry constraints.

Recent studies have explored UAV coordination and trajectory optimization for industrial IoT routing~\cite{Abou25:Centralized}, mobile edge computing~\cite{El25:Towards}, DRL-based path planning~\cite{Li22:Path}, energy-efficient resource coordination~\cite{Li25:Energy}, and covertness-aware trajectory design~\cite{Li22:Covertness}. While these works highlight the potential of optimization- and learning-based UAV control, they mainly target communication, computation, routing, or resource-management objectives rather than RSS-based emergency target localization. In contrast, the proposed rigidity-based approach is a training-free, model-based trajectory optimization framework that exploits the geometric structure of cooperative localization to improve sensing observability without requiring prior offline training.

\subsection{Applications of the Rigidity Theory}

Rigidity theory has been applied in various engineering fields, such as architectural design, molecular modeling, and robotics. 
Some studies have applied rigidity theory to agent formation maintenance \cite{Zelazo15:Decentralized, Schiano17:Bearing, Ramazani17:Rigidity} or UAV deployment \cite{Liu21:UAV}. 
However, rigidity theory has not previously been utilized in the design of UAV trajectories for effective target localization. 
Zelazo et al. \cite{Zelazo15:Decentralized} proposed a rigidity maintenance gradient controller that enables distributed agents to maintain their rigidity. 
Schiano and Giordano \cite{Schiano17:Bearing} proposed a formation strategy that maintains minimum rigidity using bearing measurements. 
Ramazani et al. \cite{Ramazani17:Rigidity} considered the formation maintenance of multi-layered agents consisting of ground and aerial vehicles operating on two planes. 
They maintained the formation of agents in the lower layer to satisfy infinitesimal rigidity \cite{Ramazani17:Rigidity}. 
Formation maintenance aims to preserve rigidity as agents move or as communication links are lost or created. 
In contrast, our study focuses on target localization. 
Our method guides the UAVs to move in directions that reduce ambiguity in the target’s position, which is fundamentally different from maintaining a predefined formation.

In \cite{Liu21:UAV}, rigidity was used for UAV aerial anchor deployment to provide localization and communication services in isolated regions through air-to-ground (A2G) channels. 
The optimization problem addressed in \cite{Liu21:UAV} aims to minimize the CRLB while ensuring rigidity among UAVs. 
However, their use of rigidity focused solely on enhancing the self-localization accuracy of the UAVs. 
By contrast, our study focuses on improving target localization performance, distinguishing it from \cite{Liu21:UAV}. 
Our optimization problem guides the UAVs to rigidify the ``framework'' that includes both the UAVs and the target, which effectively mitigates ambiguity in the target's position estimates. 
Here, a ``framework'' refers to a body-bar framework in Euclidean space comprising rigid bodies (e.g., UAVs or the target) connected by stiff bars (e.g., distance constraints). 
However, the previous study constrained rigidity only among the UAVs, which did not guarantee unique localization of the target. 
In addition, our work addresses the optimization of waypoints for moving UAVs, whereas Liu et al. \cite{Liu21:UAV} focused on deploying UAVs that hover at fixed altitudes.

\section{Problem Description} \label{sec:ProblemDescription}

We consider the following target localization problem. 
Let $\mathbf{x}_{m,t}^{\mathrm{UAV}}$ denote the position of the $m$-th UAV at time $t$, and let $\mathbf{x}_{\mathrm{tar}}$ denote the unknown position of the target. 
The UAVs collaboratively estimate the target position, denoted by $\hat{\mathbf{x}}_{\mathrm{tar}}$, using RSS measurements of the signal transmitted from the target. 
In this study, the target is assumed to remain stationary during the localization process. The UAVs are assumed to communicate with each other or with a localization server that aggregates all measurement data for centralized position estimation.
Each UAV can determine its own position using its onboard navigation system or a cooperative navigation method.

Under these assumptions, our objective is to design UAV trajectories that enable the system to satisfy the FCC emergency localization requirements (i.e., $\pm$50~m horizontal positioning error for 80\% of all 911 calls) as quickly as possible. 
To formalize this objective, we consider the following trajectory optimization problem. 
At each time step, the goal is to determine the movement direction of each UAV so that the collected measurements provide the most informative geometric configuration for target localization. 
This can be expressed as

\begin{equation}
\alpha_{m,t+1} =
\underset{\alpha}{\argmax} \;
J(\mathbf{x}_{1:M,0:t}^{\mathrm{UAV}}, \hat{\mathbf{x}}_{\mathrm{tar}}, \alpha),
\end{equation}

where $\mathbf{x}_{1:M,0:t}^{\mathrm{UAV}}$ denotes the set of UAV positions collected from time $0$ to $t$, $\alpha_{m,t+1}$ denotes the optimal movement direction of the $m$-th UAV at time $t+1$, and $J(\cdot)$ represents an objective function that quantifies how effectively the next UAV motion improves the geometric configuration for target localization and reduces the ambiguity of the target position estimate.
The specific definition of $J(\cdot)$ used in this work is introduced in Section~\ref{sec:ProposedApproach}.

\subsection{Framework Modeling} \label{subsec:GraphModeling}

\begin{figure}
    \centering
    \includegraphics[width=0.8\linewidth]{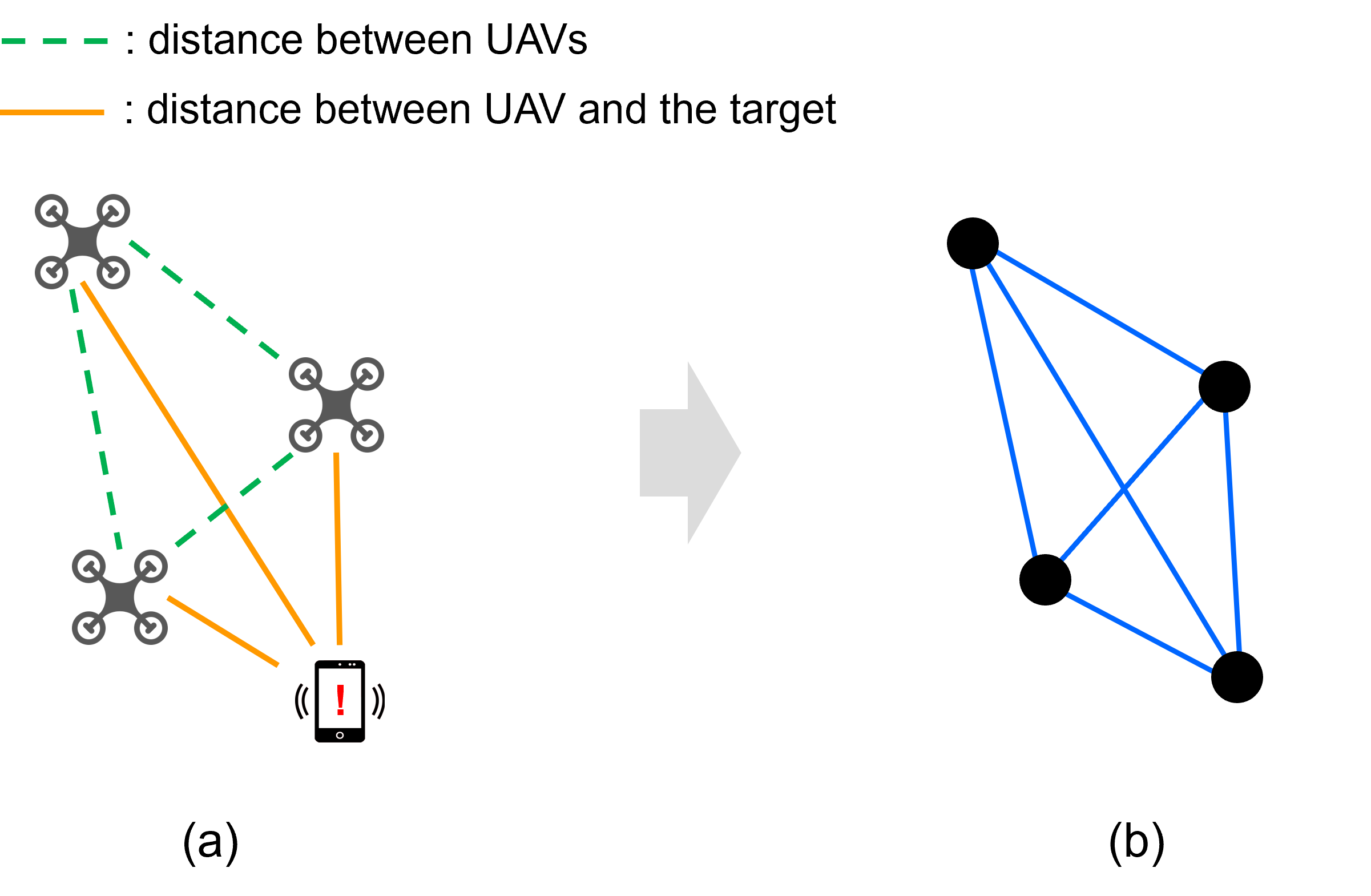}
    \caption{(a) An example of the relationship between the target and UAVs. (b) A body-bar framework model representing that relationship.}
    \label{fig:UAVFramework}
\end{figure}

The target localization problem considered in this study can be modeled as a body-bar framework (i.e., a linkage), which is a specific realization of a graph. 
A body-bar framework consists of rigid bodies connected by stiff bars \cite{Connelly13:Generic}.
Fig. \ref{fig:UAVFramework} illustrates an example target localization scenario involving three UAVs. 
The UAV-target relationship in Fig. \ref{fig:UAVFramework}(a) can be modeled as the body-bar framework shown in Fig. \ref{fig:UAVFramework}(b), where the UAVs and the target are vertices connected by distance constraints (i.e., edges with fixed lengths). 
As the UAVs move, new vertices (i.e., updated UAV positions) and edges (i.e., new distance constraints) are created in addition to the existing ones. 
The dotted green lines in Fig. \ref{fig:UAVFramework}(a) represent the distances between UAVs, which can be obtained from known UAV positions. 
The solid orange lines in Fig. \ref{fig:UAVFramework}(a) represent the distances between the UAVs and the target, which can be derived from noisy RSS measurements collected by the UAVs from the target’s signal.

The mathematical notations for the body-bar framework considered in this study are as follows. 
A body-bar framework in $\mathbb{R}^d$ is defined as a tuple $\left(G, \boldsymbol{p}\right)$, where $G = (V, E)$ is a graph and $\boldsymbol{p} : V \rightarrow \mathbb{R}^d$ is a ``configuration'' of points, with $V$ and $E$ denoting the sets of vertices and edges of the graph, respectively. 
Here, a point refers to the placement of a vertex in Euclidean space $\mathbb{R}^d$, and a configuration refers to a set of points corresponding to the vertices of $G$. 
The vertex set $V$ includes vertices representing each UAV up to time $t$ (e.g., the vertex of the $m$-th UAV at time $t$ is denoted as $v^{\mathrm{UAV}}_{m,t}$), as well as a vertex representing the stationary target, $v_\mathrm{tar}$. 
The configuration $\boldsymbol{p}$ maps the vertices to points in $\mathbb{R}^d$, meaning each vertex is associated with specific coordinates in $d$-dimensional space. 
In addition, the length of an edge $uv \in E$ is defined as the Euclidean distance between two points $\boldsymbol{p}(u)$ and $\boldsymbol{p}(v)$. 
Here, $u$ and $v$ are arbitrary vertices in $V$ (i.e., $u, v \in V$), and $uv$ denotes an edge connecting them. 
$\boldsymbol{p}(u)$ and $\boldsymbol{p}(v)$ represent the corresponding points in Euclidean space $\mathbb{R}^d$ for vertices $u$ and $v$, respectively.

\subsection{Target Position Estimation} \label{subsec:MeasurementModeling}

As each UAV knows its own position using onboard navigation sensors, the configuration of $v^{\mathrm{UAV}}_{m,t}$ is expressed as $\boldsymbol{p}(v^\mathrm{UAV}_{m,t}) = \mathbf{x}^\mathrm{UAV}_{m, t}$, where $\mathbf{x}^\mathrm{UAV}_{m,t}$ denotes the position of the $m$-th UAV at time $t$.
However, the position of the target (i.e., $\boldsymbol{p}(v_\mathrm{tar})$) is unknown and must be estimated from RSS measurements.

The RSS measured by the $m$-th UAV at time $t$, $\hat{P}_{m, t}$, is modeled using a log-distance path loss model as follows \cite{Uluskan20:Noncausal, Koohifar16:Receding, Vaghefi12:Cooperative}:
\begin{equation} \label{eq:Path_loss_model}
\begin{split}
\hat{P}_{m, t} &= P_{m,t} + n_{m, t} = P_{0} - 10 \beta \log_{10} \frac{d_{m, t}}{d_0} + n_{m, t},\\
d_{m, t} &= \|\mathbf{x}^\mathrm{UAV}_{m,t}-\mathbf{x}_\mathrm{tar}\|, \\
n_{m, t} &\sim \mathcal{N}(0, \sigma_\mathrm{dB}^2),
\end{split}
\end{equation}
where $P_0$ (in dBm) is the RSS at the reference distance $d_0$ from the target; $d_0$ is a reference distance (set to 2.5 m in this study without loss of generality); $\beta$ is the path loss exponent; $n_{m, t}$ represents the log-normal shadowing, modeled as a normal distribution in the dB scale with standard deviation $\sigma_\mathrm{dB}$; $\mathbf{x}_\mathrm{tar} = \trp{ \left[ x, y \right] }$ is the true position of the target; and $\|\cdot\|$ denotes the Euclidean norm. 
As in previous studies \cite{Uluskan20:Noncausal, Koohifar16:Receding}, the values of $\beta$ and $P_0$ are assumed to be known. 
Otherwise, they can be estimated using the methods described in \cite{Vaghefi12:Cooperative, Wang12:Received}.

When $M$ UAVs over a time horizon of $T$ are considered, the position of the target $\boldsymbol{p}(v_\mathrm{tar})$ can be estimated using maximum likelihood estimation (MLE) as follows:
\begin{equation} \label{eq:MLE}
    \begin{split}
    \boldsymbol{p}(v_\mathrm{tar}) &= \hat{\mathbf{x}}_\mathrm{tar} \\
    &= \argmin_{\mathbf{x}_\mathrm{tar}} \, \sum_{t=0}^{T} \sum_{m=1}^{M} \left( \hat{P}_{m,t} - P_{0} + 10 \beta \log_{10} \frac{d_{m,t}}{d_0} \right)^2.
    \end{split}
\end{equation}
To obtain an approximate solution to the maximum likelihood (ML) problem in (\ref{eq:MLE}), various methods can be employed, including iterative algorithms \cite{Kulkarni10:Particle, Lee23:Performance_Comparison}, convex relaxation \cite{Vaghefi12:Cooperative, Ouyang10:Received}, and linear approximation \cite{Vaghefi12:Cooperative}.

However, developing improved search or approximation methods for solving ML problems is beyond the scope of this study. 
The objective of this study is to reduce the time required to localize a target to an acceptable level in emergency scenarios by efficiently guiding UAVs to minimize localization error. 
To achieve this, we focus on rapidly reducing geometric ambiguity in the target position estimate by utilizing the concept of rigidity.

The position ambiguities shown in Fig. \ref{fig:PositionAmbiguity} refer to the presence of multiple local minima with similar costs in the cost function described in (\ref{eq:MLE}). 
In other words, there may be several solutions with comparable costs, making it difficult to accurately determine the true target position. 
Furthermore, due to errors in RSS measurements, some candidate solutions may yield lower costs than the true solution. 
Given these considerations, it is important to note that estimation methods---regardless of their specific type or implementation---may result in significant localization errors if position ambiguity is not properly resolved.

In the next section, we briefly introduce the concept of rigidity and explain how the rigidity of frameworks can be determined. 
The theorems presented in Section \ref{sec:RigidityTheoryPreliminaries} are used to design the rigidity-based UAV trajectory optimization problem proposed in Section \ref{sec:ProposedApproach}.

\section{Rigidity Theory Preliminaries}
\label{sec:RigidityTheoryPreliminaries}

This section reviews the fundamental concepts of rigidity theory as applied in this study for UAV trajectory optimization.
We first introduce the definitions of rigidity (also referred to as local rigidity) and global rigidity, followed by theorems used to verify the rigidity and global rigidity of a given framework. 
For more details on rigidity theory, see \cite{Sitharam18:Handbook}.

\subsection{Definitions}

The definition of global rigidity begins with the concepts of equivalence and congruence of frameworks. 
Two frameworks in $\mathbb{R}^d$, $(G, \boldsymbol{p})$ and $(G, \boldsymbol{q})$, are \textit{equivalent} if the lengths of all corresponding edges are the same \cite{Sitharam18:Handbook}. 
They are \textit{congruent} if all distances between corresponding points are equal, i.e., $\|\boldsymbol{p}(u) - \boldsymbol{p}(v)\| = \|\boldsymbol{q}(u) - \boldsymbol{q}(v)\|$ for all $u, v \in V$ \cite{Sitharam18:Handbook}.

\vspace{1em}
\begin{definition}[Global Rigidity \cite{Sitharam18:Handbook}]
A body-bar framework $(G, \boldsymbol{p})$ in $\mathbb{R}^d$ is globally rigid if every body-bar framework that is equivalent to $(G, \boldsymbol{p})$ is also congruent to $(G, \boldsymbol{p})$.
\end{definition}
\vspace{1em}

\begin{definition}[Local Rigidity \cite{Sitharam18:Handbook}] \label{def:rigidity}
A body-bar framework $(G, \boldsymbol{p})$ in $\mathbb{R}^d$ is rigid if there exists $\varepsilon > 0$ such that every body-bar framework $(G, \boldsymbol{q})$ that is equivalent to $(G, \boldsymbol{p})$ and satisfies $\|\boldsymbol{p}(v) - \boldsymbol{q}(v)\| < \varepsilon$ for all $v \in V$ is congruent to $(G, \boldsymbol{p})$.
\end{definition}
\vspace{1em}

\begin{figure}
    \centering
    \includegraphics[width=0.75\linewidth]{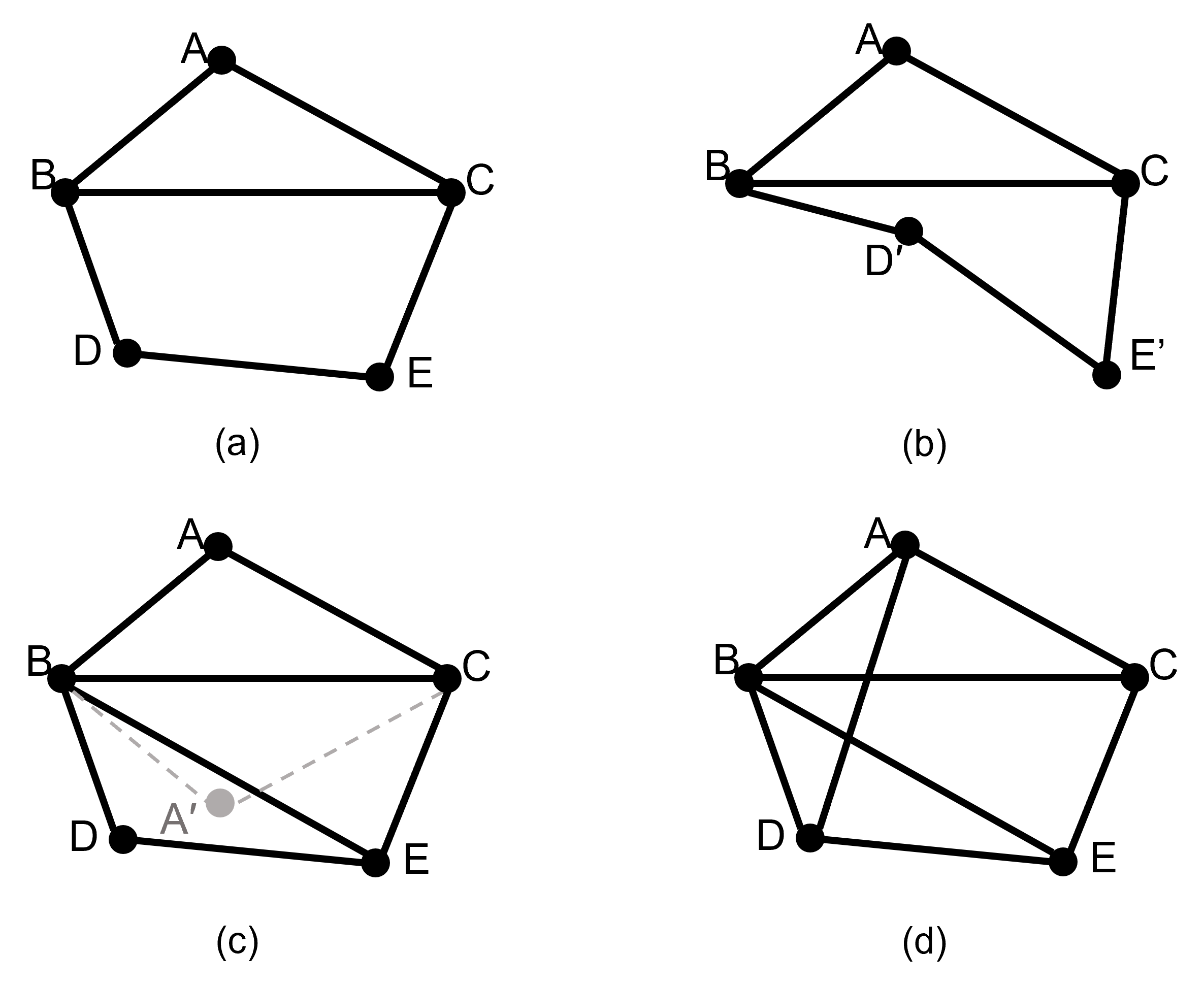}
    \caption{Examples of nonrigid, rigid, and globally rigid frameworks in $\mathbb{R}^2$. 
    (a) A nonrigid framework that can undergo smooth deformation, as illustrated in (b). 
    (c) A rigid but not globally rigid framework. Point A can be flipped to a symmetric point A$'$ across edge BC while maintaining the lengths of all edges. 
    (d) A globally rigid framework.}
    \label{fig:frameworks_example}
\end{figure}

Fig. \ref{fig:frameworks_example} shows examples of nonrigid, rigid, and globally rigid frameworks. 
In Definition \ref{def:rigidity}, rigidity implies that there is no ``smooth deformation'' of the framework $(G, \boldsymbol{p})$ that preserves the lengths of all edges. 
However, some discontinuous deformations (e.g., flipping or discontinuous flexing) can occur in rigid frameworks that are not globally rigid. 
Fig. \ref{fig:frameworks_example}(c) illustrates an example of flip ambiguity: even if the positions of all other points from B to E are known, the position of point A remains ambiguous (i.e., it can be either A or A$'$). 
If the framework $(G, \boldsymbol{p})$ is globally rigid, it represents a unique realization of the graph $G$. 
Therefore, a point in the framework is uniquely localizable if the positions of the other points in $\mathbb{R}^d$ are known.

\subsection{Theorems for Checking Rigidity}

Determining the rigidity of frameworks for $d \geq 2$ and the global rigidity for $d \geq 1$ is known to be NP-hard \cite{Saxe79:Embeddability, Connelly13:Generic}. 
However, when frameworks are restricted to be \textit{generic}, it is possible to determine their generic rigidity and generic global rigidity in polynomial time. 
A framework is considered generic if the coordinates of all its points are algebraically independent \cite{Sitharam18:Handbook}. 
The rigidity matrix theorem can be used to determine the generic rigidity of any body-bar framework.

\vspace{1em}
\begin{definition}[Rigidity Matrix \cite{Sitharam18:Handbook}]
The rigidity matrix $\mathbf{R}(G, \boldsymbol{p})$ of the framework $(G, \boldsymbol{p})$ in $\mathbb{R}^d$ is an $|E| \times d|V|$ matrix, where the rows are indexed by the edges. 
Here, $|E|$ and $|V|$ denote the number of edges and vertices, respectively. 
The row of the rigidity matrix corresponding to the edge $uv \in E$, which connects a pair of vertices $u, v \in V$, is given by:
\begin{equation}
\begin{aligned}
\mathbf{R}(G, \boldsymbol{p})_{uv} &= \left(
\begin{array}{ccccccccccc}
0 & \hspace{-0.5em} \cdots & \hspace{-0.5em} 0 & \boldsymbol{p}(u) - \boldsymbol{p}(v) & 0 & \hspace{-0.5em} \cdots & \hspace{-0.5em} 0
\end{array}
\right. \\
&\left.
\begin{array}{ccccccccccc}
\hspace{3.5em} 0 & \hspace{-0.5em} \cdots & \hspace{-0.5em} 0 & \boldsymbol{p}(v) - \boldsymbol{p}(u) & 0 & \hspace{-0.5em} \cdots & \hspace{-0.5em} 0
\end{array}
\right).
\end{aligned}
\end{equation}
Each row of the rigidity matrix contains $d|V|$ columns. 
Among these, the columns corresponding to vertices $u$ and $v$ contain the $d$ coordinates of $(\boldsymbol{p}(u) - \boldsymbol{p}(v))$ and $(\boldsymbol{p}(v) - \boldsymbol{p}(u))$, respectively, while all remaining columns are zero.
\end{definition}
\vspace{1em}

\begin{theorem}[Rigidity Matrix Theorem \cite{Sitharam18:Handbook}]
\raggedright
A body-bar framework $(G,\boldsymbol{p})$ in $\mathbb{R}^d$ is generically rigid if and only if $\mathrm{rank}(\mathbf{R}(G,\boldsymbol{p}))=d|V|-\frac{d(d+1)}{2}$.
\end{theorem}
\vspace{1em}

Connelly et al. \cite{Connelly13:Generic} proved that a special class of frameworks, known as body-bar frameworks, is generically globally rigid in $\mathbb{R}^d$ for any $d \geq 1$ if and only if it is generically redundantly rigid. 
A redundantly rigid framework remains rigid even after the removal of any single edge \cite{Connelly13:Generic}.

\begin{figure}
    \centering
    \includegraphics[width=0.85\linewidth]{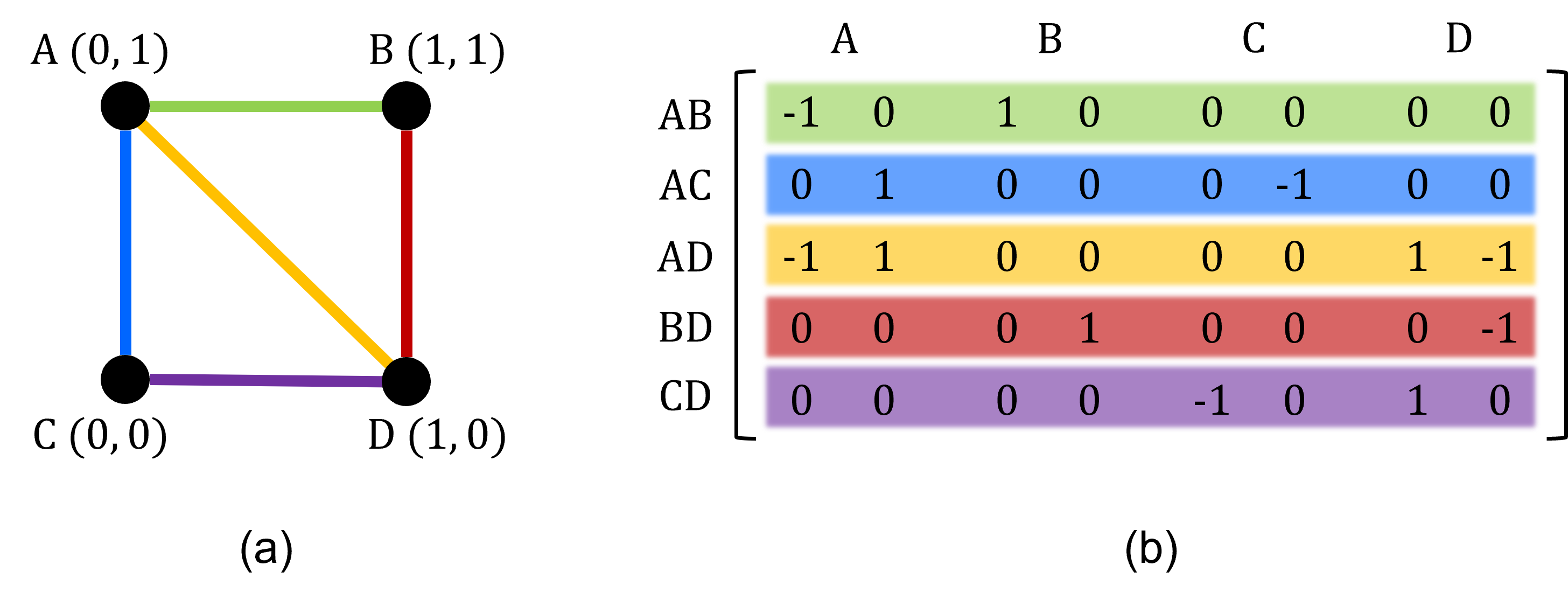}
    \caption{Example of constructing a rigidity matrix. (a) A body-bar framework in $\mathbb{R}^2$. (b) The corresponding rigidity matrix, with each row representing an edge constraint in the framework.}
    \label{fig:RigidityMatrix}
\end{figure}

Fig. \ref{fig:RigidityMatrix} shows an example of a rigidity matrix in $\mathbb{R}^2$. 
For the framework shown in Fig. \ref{fig:RigidityMatrix}(a), the corresponding rigidity matrix is constructed as illustrated in Fig. \ref{fig:RigidityMatrix}(b). 
The rigidity matrix in Fig. \ref{fig:RigidityMatrix}(b) consists of 5 rows and 8 columns, corresponding to the number of edges and twice the number of vertices, respectively.

According to the rigidity matrix theorem, a framework in $\mathbb{R}^2$ is generically rigid if its rigidity matrix has rank $2|V| - 3$. 
The framework in Fig. \ref{fig:RigidityMatrix}(a) is generically rigid because the rank of the rigidity matrix in Fig. \ref{fig:RigidityMatrix}(b) is 5, which equals $2|V| - 3$ for a framework with four vertices. 
However, the framework is not generically globally rigid because it is not redundantly rigid; it becomes flexible when edge $AD$ is removed.

\section{Proposed Approach}
\label{sec:ProposedApproach}

In the previous section, theorems for determining rigidity and global rigidity were presented. 
The main idea of our approach is to optimize the trajectories of the UAVs by increasing the rigidity of the framework formed by the UAVs and the target, thereby reducing the position ambiguity of the target. 
If a framework composed of UAVs and a target is nonrigid, certain deformations can be applied to the framework. 
This implies that multiple candidate positions in Euclidean space may satisfy the same set of distance constraints, as illustrated by examples D and D$'$ in Figs. \ref{fig:frameworks_example}(a) and \ref{fig:frameworks_example}(b), and A and A$'$ in Fig. \ref{fig:frameworks_example}(c). 
Thus, if the framework is nonrigid, the target may lie at positions other than the one estimated by the MLE, even when the UAV positions are fixed.

Conversely, if the framework is globally rigid, no deformation can be applied. 
This implies that the target's position can be uniquely determined. 
Therefore, by increasing the rigidity of the framework, we can better ensure the uniqueness of the target's position, thereby reducing position ambiguity.

\subsection{Rigidity Matrix Formulation}

In this study, we construct the rigidity matrix based on the previous positions of UAVs from time 0 to $t$, the candidate position of the $m$-th UAV at time $t+1$, and the estimated target position obtained by the MLE in (\ref{eq:MLE}).
For example, assuming $t = 0$ and two UAVs, the rigidity matrix for determining the optimal movement direction of the first UAV at $t = 1$ is given in (\ref{eq:RigidityMatrix}).

\begin{table*}
\begin{equation}\label{eq:RigidityMatrix}
\begin{aligned}
\mathbf{R}(G,\boldsymbol{p}) &= \\
&\scalebox{1.0}{$\left(\hspace{0.1em}
\begin{array}{@{}cccc@{}}
\vspace{-0.6em}\\
\vspace{1em}
\boldsymbol{p}(v^\text{UAV}_{1,0})-\boldsymbol{p}(v^\text{UAV}_{2,0}) & \boldsymbol{p}(v^\text{UAV}_{2,0})-\boldsymbol{p}(v^\text{UAV}_{1,0}) & 0 \hspace{3em} 0 & 0 \hspace{3em} 0 \\
\vspace{1em}
\boldsymbol{p}(v^\text{UAV}_{1,0})-\boldsymbol{p}(v^\text{UAV}_{1,1}) & 0 \hspace{3em} 0 & \boldsymbol{p}(v^\text{UAV}_{1,1})-\boldsymbol{p}(v^\text{UAV}_{1,0}) & 0 \hspace{3em} 0 \\
\vspace{1em}
\boldsymbol{p}(v^\text{UAV}_{1,0})-\boldsymbol{p}(v_\text{tar}) & 0 \hspace{3em} 0 & 0 \hspace{3em} 0 & \boldsymbol{p}(v_\text{tar})-\boldsymbol{p}(v^\text{UAV}_{1,0}) \\
\vspace{1em}
0 \hspace{3em} 0 & \boldsymbol{p}(v^\text{UAV}_{2,0})-\boldsymbol{p}(v^\text{UAV}_{1,1}) & \boldsymbol{p}(v^\text{UAV}_{1,1})-\boldsymbol{p}(v^\text{UAV}_{2,0}) & 0 \hspace{3em} 0 \\
\vspace{1em}
0 \hspace{3em} 0 & \boldsymbol{p}(v^\text{UAV}_{2,0})-\boldsymbol{p}(v_\text{tar}) & 0 \hspace{3em} 0 & \boldsymbol{p}(v_\text{tar})-\boldsymbol{p}(v^\text{UAV}_{2,0}) \\
\vspace{0.5em}
0 \hspace{3em} 0 & 0 \hspace{3em} 0 & \boldsymbol{p}(v^\text{UAV}_{1,1})-\boldsymbol{p}(v_\text{tar}) & \boldsymbol{p}(v_\text{tar})-\boldsymbol{p}(v^\text{UAV}_{1,1}) \\
\end{array}
\hspace{0.1em}\right)$}.
\end{aligned}
\end{equation}
\end{table*}

As UAVs move over time, the size of the rigidity matrix increases accordingly.
Since the computational cost of SVD grows with the matrix size, using all UAV vertices from time 0 to $t$ quickly becomes inefficient.
To address this issue, Section \ref{subsec:PriorityQueue} proposes a method that maintains the rigidity matrix at a constant size by selectively using only high-priority UAV vertices via a priority queue, thereby enabling real-time implementation.

\subsection{Singular Value Decomposition}

The key to formulating the optimization problem in this study lies in the rank of the rigidity matrix, which must be equal to $d|V| - \frac{d(d+1)}{2}$ if the given framework is generically rigid. 
To check the rank of the rigidity matrix $\mathbf{R}(G, \boldsymbol{p})$, we perform SVD as follows:
\begin{equation} \label{eq:SVD}
\mathbf{R}(G, \boldsymbol{p}) = \mathbf{U} \mathbf{\Sigma} \mathbf{B}^\mathrm{T},
\end{equation}
where $\mathbf{U}$ contains the left singular vectors, $\mathbf{\Sigma}$ is a rectangular diagonal matrix whose diagonal entries are the singular values of $\mathbf{R}(G, \boldsymbol{p})$ sorted in descending order, and $\mathbf{B}$ contains the right singular vectors.

According to the rigidity matrix theorem, if the framework is generically rigid, we obtain $d|V| - \frac{d(d+1)}{2}$ nonzero singular values. 
This implies that all other singular values, except for the first $d|V| - \frac{d(d+1)}{2}$ values of the matrix $\mathbf{\Sigma}$, are zero. 
For example, when $d = 2$, a generically rigid framework is expected to yield $2|V| - 3$ nonzero singular values. 
In other words, if the framework is flexible, the $(2|V| - 3)$-th singular value (i.e., $\sigma_{2|V|-3}$) must be zero. 
Therefore, the more rigid the framework, the greater the value of $\sigma_{2|V|-3}$ is expected to be.

It should be noted that the proposed method does not use the rank of the rigidity matrix itself as the trajectory optimization metric. For a fully connected generic framework in $\mathbb{R}^2$, the rigidity matrix already has rank $2|V|-3$. However, even among such rigid configurations, the smallest nonzero singular value can vary significantly depending on the geometric arrangement of the UAVs. In particular, near-collinear or poorly distributed UAV configurations yield much smaller values of $\sigma_{2|V|-3}$, indicating that the sensing geometry is poorly conditioned for localization. In contrast, well-spread configurations produce larger values of $\sigma_{2|V|-3}$ and provide more informative geometric diversity. Therefore, the proposed trajectory optimization seeks to maximize $\sigma_{2|V|-3}$ rather than relying on the rank itself.
This metric differs from FIM-based criteria in that it is not a scalarization of the estimator covariance at the current operating point. Instead, it measures how well the overall UAV--target framework is geometrically conditioned, thereby guiding UAVs toward configurations that improve target distinguishability even when the current MLE is biased or ambiguous.

However, $\sigma_{2|V|-3}$ should be interpreted as a local/geometric conditioning metric of the estimated UAV--target framework. Increasing $\sigma_{2|V|-3}$ makes the framework less susceptible to infinitesimal geometric degeneracy and improves target distinguishability, but it does not mathematically guarantee the elimination of flip ambiguity or all local minima of the RSS likelihood function. Therefore, the proposed objective mitigates position ambiguity by encouraging well-conditioned sensing geometries, rather than eliminating all ambiguous target solutions.

\subsection{Rigidity-Based Trajectory Optimization Problem Design}

Given the RSS measurements obtained up to time $t$ and the known positions of all UAVs until time $t$, our method optimizes the movement direction of the $m$-th UAV at time $t+1$ (denoted as $\alpha_{m,t+1}$). 
By guiding the UAVs in the direction that increases $\sigma_{2|V|-3}$, they can collect RSS measurements that help reduce the position ambiguity of the target. 
Therefore, we formulate a UAV trajectory optimization problem that maximizes $\sigma_{2|V|-3}$ at $t+1$ (i.e., $\sigma_{2|V|-3, t+1}$) as follows:
\begin{equation} \label{eq:ProposedOptimization}
\alpha_{m,t+1} = \underset{\alpha}{\argmax} \; \sigma_{2|V|-3, t+1}(\alpha),
\end{equation}
where $\alpha_{m,t+1}$ denotes the optimal movement direction for the $m$-th UAV at time $t+1$.

The position of the \textit{m}-th UAV in the next epoch $t+1$ is given as a function of $\alpha$ as follows:
\begin{equation}
\begin{split}
x^\mathrm{UAV}_{m,t+1} \left( \alpha \right) &= x^\mathrm{UAV}_{m,t} + l \cdot \mathrm{cos} \left( \alpha \right)\\[0.6em]
y^\mathrm{UAV}_{m,t+1} \left( \alpha \right) &= y^\mathrm{UAV}_{m,t} + l \cdot \mathrm{sin} \left( \alpha \right),
\end{split}
\end{equation}
where $l$ is the separation between two subsequent waypoints of the UAV.

In summary, the overall flow of the proposed rigidity-based UAV trajectory optimization is outlined in Algorithm \ref{alg:trajectoryOptimization}. 
After constructing the rigidity matrix and computing the singular values, the optimal movement angle for the $m$-th UAV at time $t+1$ is determined.

\setlength{\algomargin}{0pt}
\begin{algorithm} [tbp]
\caption{Rigidity-based UAV trajectory optimization} \label{alg:trajectoryOptimization}
\vspace{0.2em}
\KwData{Positions of all UAVs until $t$, estimated target position, candidate moving angles $A$, separation $l$}
\KwResult{Optimal moving angle of the $m$-th UAV $\alpha_{m,t+1}$}

\BlankLine

$\sigma_\text{opt} \gets -\text{Inf}$\;
\vspace{0.2em}
\For{every $\alpha \in A$}{
    \vspace{0.3em}
    $x^\mathrm{UAV}_{m,t+1} \gets x^\mathrm{UAV}_{m,t} + l \cdot \mathrm{cos} \left( \alpha \right)$\;
    \vspace{0.4em}
    $y^\mathrm{UAV}_{m,t+1} \gets y^\mathrm{UAV}_{m,t} + l \cdot \mathrm{sin} \left( \alpha \right)$\;
    \vspace{0.3em}
    Construct the rigidity matrix considering the positions of UAVs from 0 to $t$, $x^\mathrm{UAV}_{m,t+1}$, $y^\mathrm{UAV}_{m,t+1}$, and the estimated target position\;
    \vspace{0.1em}
    Calculate the singular value $\sigma_{2|V|-3, t+1}$\;
\vspace{0.2em}
    \If{$\sigma_{2|V|-3, t+1} > \sigma_\text{opt}$}{
    \vspace{0.2em}
        $\sigma_\text{opt} \gets \sigma_{2|V|-3, t+1}$\;
        $\alpha_{m,t+1} \gets \alpha$\;
    \vspace{0.15em}
    }
}
\end{algorithm}

\subsection{Rigidity Matrix Reduction} \label{subsec:PriorityQueue}

The computational complexity of computing the singular value decomposition (SVD) depends on the size of the matrix. 
For example, the \texttt{svd} function in MATLAB has a computational complexity of $\mathcal{O}(\max(m,n)\min(m,n)^2)$ for an $m \times n$ matrix \cite{Wang15:A}.
As the size of the rigidity matrix increases over time, SVD computation becomes increasingly expensive. 
To enable real-time applications, we develop a method to reduce the size of the rigidity matrix by eliminating less informative vertices.

When eliminating vertices, we prioritized UAV vertices that played a more significant role in enhancing the rigidity of the framework---specifically, those that contributed to increasing $\sigma_{2|V|-3}$. 
To achieve this, we constructed a priority queue in which elements were sorted based on the change in $\sigma_{2|V|-3}$ values at each epoch (i.e., $\sigma_{2|V|-3, t} - \sigma_{2|V|-3, t-1}$ for $t > 1$). 
Elements with greater change were assigned higher priority. 
When the desired number of UAV vertices is $K$, the priority queue retains only the $K$ elements with the highest priority, and all other UAV vertices are discarded.

The algorithm for maintaining the priority queue is detailed in Algorithm \ref{alg:priorityQueue}. 
The computational complexities of inserting and removing elements from the priority queue are both $\mathcal{O}(\log n)$ when implemented using a binary heap, where $n$ is the number of elements in the queue.

\begin{algorithm} [tbp]
\caption{Procedure for rigidity matrix size reduction}
\label{alg:priorityQueue}
\KwData{Total epochs $T$, desired number of UAV vertices $K$}
\KwResult{$selectedVertices$}

\BlankLine
\textbf{State:} 1-indexed array $H$ of tuples $(\Delta\sigma, v)$ maintained as a \emph{min-heap} by key $\Delta\sigma$

\SetKwFunction{Insert}{Insert}
\SetKwFunction{ExtractMin}{ExtractMin}
\SetKwFunction{MinHeapify}{MinHeapify}

\BlankLine
\textbf{Procedure } \Insert{$(\Delta\sigma, v)$}:

append $(\Delta\sigma, v)$ to $H$; $i \gets |H|$\;
\While{$i>1$ \textbf{and} $H[i].\Delta\sigma < H[\lfloor i/2 \rfloor].\Delta\sigma$}{
  swap $H[i]$ and $H[\lfloor i/2 \rfloor]$; $i \gets \lfloor i/2 \rfloor$\;
}

\BlankLine
\textbf{Procedure } \ExtractMin{}:

$min \gets H[1]$\;
$H[1] \gets H[|H|]$; remove last element\;
\MinHeapify{$1$}\;

\BlankLine
\textbf{Procedure } \MinHeapify{$i$}:

\While{$2i \le |H|$}{
  $j \gets 2i$ \tcp*{left child}
  \If{$j+1 \le |H|$ \textbf{and} $H[j+1].\Delta\sigma < H[j].\Delta\sigma$}{$j \gets j+1$ \tcp*{pick smaller child}}
  \If{$H[i].\Delta\sigma \le H[j].\Delta\sigma$}{\textbf{break}}
  swap $H[i]$ and $H[j]$; $i \gets j$\;
}

\BlankLine
\textbf{Main:}
\BlankLine
initialize empty min-heap $H$\;
\For{$t=1$ \KwTo $T$}{
  new vertex $v^\text{UAV}_t$ appears\;
  $\Delta\sigma_t \gets \sigma_{2|V|-3, t} - \sigma_{2|V|-3, t-1}$\;

  \eIf{$|H| < K$}{
    \Insert{$(\Delta\sigma_t, v^\text{UAV}_t)$}\;
  }{
    \If{$\Delta\sigma_t > H[1].\Delta\sigma$}{
      \ExtractMin{}\;
      \Insert{$(\Delta\sigma_t, v^\text{UAV}_t)$}\;
    }
  }
}
$selectedVertices \gets$ vertices stored in $H$\;

\end{algorithm}

\section{Simulation Settings} \label{sec:SimulationSettings}

\subsection{Simulation Scenario}
We evaluated the target localization performance of the proposed rigidity-based UAV trajectory optimization method under the following mission scenario. 
UAVs are initially dispatched to a coarse position estimate of the target, obtained via cellular-based localization, and begin their search operation accordingly. 
According to the Korea Communications Commission, the average localization accuracy of cellular network-based positioning ranges from 107.0 m to 146.3 m \cite{KCC23}.

Based on this range, it is assumed that the UAVs begin the mission at the base location of $(-125\,\mathrm{m}, -125\,\mathrm{m})$, while the true target is located at the origin $(0, 0)$. 
The RSS at the reference distance, $P_0$, is set to 3 dBm, and the path loss exponent $\beta$ is set to 2.

We assumed a shadowing standard deviation $\sigma_\mathrm{dB}$ of 5 dB, providing a 1--2 dB margin compared to the ITU-R path loss model. 
Specifically, a 3 dB standard deviation is recommended in the ITU-R Urban Micro LOS path loss model, while values of 4 dB are recommended for Urban Macro, Suburban Macro, and Rural Macro LOS models at the distances considered in this study. 
This shadow fading represents long-term variations in signal attenuation due to environmental obstacles such as buildings and trees. 
Accordingly, the simulation environments are designed to reflect realistic conditions, enhancing the practical relevance of the performance evaluation.

The UAV speed was set to 5 m/s, meaning that $l$ was set to 5 m and each epoch corresponds to 1 second. 
This setup is consistent with the parameters used in prior work \cite{Uluskan20:Noncausal}, where UAV movement of five units per epoch was also assumed. 
Note that the proposed method is not limited to this specific speed and can be extended to UAVs operating at different velocities.

To solve the localization problem in (\ref{eq:MLE}), we perform a uniform grid search over a square domain centered at the origin with side length 750 m.
The grid step size is 2.5 m; candidate positions are sampled every 2.5 m along both axes.
The value of $K$ was fixed at 20 for all simulations, except in Section \ref{subsec:CodeExecutionTimeAnalysis}, where execution time was analyzed for varying values of $K$.

To solve the UAV trajectory optimization problem, candidate movement directions were discretized at $1^{\circ}$ intervals.  
To prevent sharp turns, the allowable movement direction at each time step was constrained to lie within $\pm 20^{\circ}$ of the previous movement direction.
For all simulation cases, we conducted 1,000 Monte Carlo iterations to ensure statistical reliability in the performance evaluation.

\subsection{Comparative Algorithms and Figures of Merit} \label{subsec:ComparisonAndMetrics}

We compared the proposed rigidity-based UAV trajectory optimization method with several existing variants of FIM-based methods, as well as a straight-line strategy, as described below:
\begin{enumerate}
    \item \textit{D-optimality:} This criterion is the most widely used in prior studies \cite{Wang18:Performance, Uluskan20:Noncausal, Koohifar16:Receding, LeCadre97:Discrete, Passerieux98:Optimal, He19:The, He19:Trajectory}. The UAV's movement direction is optimized to maximize the determinant of the FIM, which is related to the volume of the uncertainty ellipsoid.

    \item \textit{A-optimality:} The UAV's movement direction is optimized to minimize the trace of the inverse of the FIM \cite{Shahidian16:Path, Sahu22:Optimal}, which corresponds to the average variance of the parameter estimates.

    \item \textit{E-optimality:} The UAV's movement direction is optimized to maximize the minimum eigenvalue of the FIM \cite{Shahidian16:Path, Sahu22:Optimal}, which corresponds to minimizing the largest axis of the uncertainty ellipsoid.

    \item \textit{Noncausal:} The UAV's movement direction is optimized by considering not only previously visited waypoints but also the remaining linear future trajectory. This approach was proposed in \cite{Uluskan20:Noncausal}.

    \item \textit{Straight Line:} The UAV moves directly toward the estimated position of the target at each time step.
\end{enumerate}

We evaluated the performance of each approach using the following figures of merit:
\begin{enumerate}
    \item \textit{Success Rate (\%):} The percentage of simulation runs that achieved a localization error of less than 50 m. This threshold is based on the FCC requirement for horizontal positioning accuracy, which mandates that 80\% of 911 calls must fall within a 50 m error.

    \item \textit{Search Time (s):} The minimum duration required for the UAVs to localize at least 80\% of 911 calls with an accuracy of 50 m. In other words, it corresponds to the time at which the success rate exceeds 80\%.

    \item \textit{Root Mean Squared Error (RMSE) (m):} The root mean squared error of the target localization, computed across all Monte Carlo simulation runs.

    \item \textit{Code Execution Time (s):} The central processing unit (CPU) time required for a UAV to compute its next movement direction. This metric is evaluated in detail in Section \ref{subsec:CodeExecutionTimeAnalysis}.
\end{enumerate}

\section{Simulation Results} \label{sec:SimulationResults}

\subsection{Performance Comparison with Existing Methods} \label{subsec:PerformanceComparison}

\begin{figure} 
    \centering
    \includegraphics[width=0.7\linewidth]{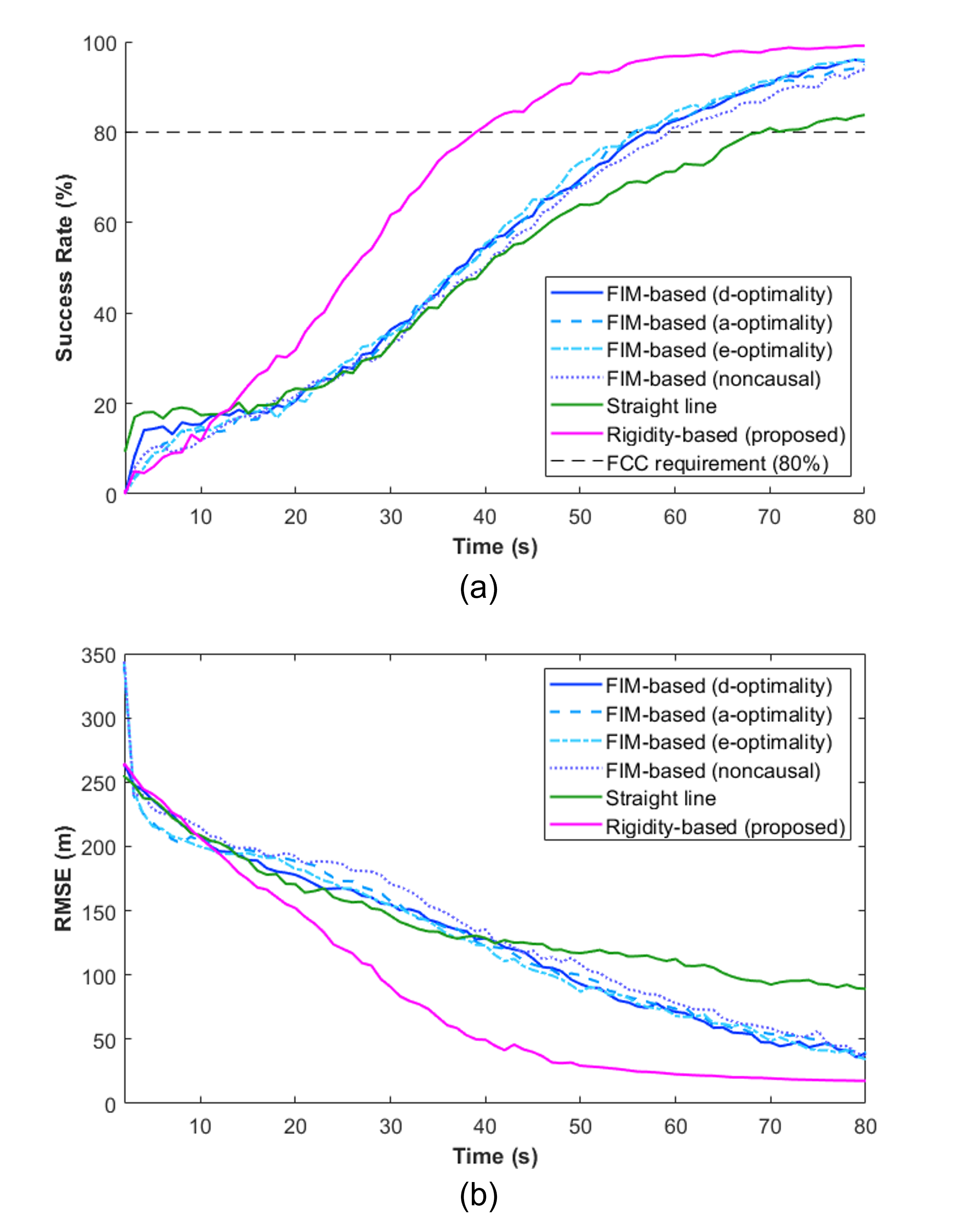}
    \caption{Performance comparison. (a) Success rate over time, measured as the percentage of simulation iterations that achieved localization within a 50~m error. (b) Localization RMSE over time.}
    \label{fig:Ideal_SuccessRate}
\end{figure}

This subsection presents a performance comparison between the proposed rigidity-based UAV trajectory optimization method and existing FIM-based approaches for target localization. 
The results for the two-UAV scenario are summarized as follows. 
Figs. \ref{fig:Ideal_SuccessRate}(a) and \ref{fig:Ideal_SuccessRate}(b) show the success rate and RMSE over time, respectively, for all methods. 
As illustrated in Fig. \ref{fig:Ideal_SuccessRate}(b), the proposed rigidity-based method achieves faster RMSE reduction than the FIM-based methods, indicating more efficient convergence in target localization.

Table \ref{tab:Ideal_Time} presents the search time required for each method to satisfy the FCC criterion. 
The rigidity-based approach achieves an 80\% success rate within 39.0 seconds, representing a 32.9\% reduction in time compared to the FIM-based method using D-optimality. 
Notably, even when the success rate threshold is increased to 90\%, the proposed method continues to outperform all FIM-based approaches.

\begin{table} [bp]
\centering
\small
\caption{Search Time Required to Meet FCC Localization Requirements for the Proposed and Baseline Approaches}
\label{tab:Ideal_Time}
{\renewcommand{\arraystretch}{1.6}
 \begin{tabular}{M{1.3cm} | M{0.9cm} | M{2.0cm} | M{2.6cm}}
 \Xhline{1.0pt}
 \multicolumn{2}{c|}
 {\textbf{Requirements}} & \multicolumn{2}{c}{\textbf{Time Spent to Meet Requirements (s)}} \\
 \hline
 \vspace{3pt}\textbf{Accuracy}\vspace{3pt} & \vspace{3pt}\textbf{Rate}\vspace{3pt} & \vspace{2pt}\makecell{\textbf{FIM-based}\\\textbf{(D-optimality)}}\vspace{2pt} & \vspace{2pt}\makecell{\textbf{Rigidity-based}\\\textbf{(proposed)}}\vspace{2pt} \\
\hline
\multirow{2}{*}{50 m} 
 & 80\% & 58.1 & 39.0 \\
 & 90\% & 68.8 & 47.6 \\
 \Xhline{1.0pt}
\end{tabular}}
\end{table}

Among the FIM-based variants, only marginal performance differences were observed regardless of the optimality criterion (e.g., D-, A-, or E-optimality), and all variants consistently underperformed the proposed method. This indicates that the primary performance bottleneck is not the choice of scalarization, but the shared reliance on an operating point that may be unreliable during the early stages of the mission. As discussed in Section~\ref{sec:Introduction}, when measurement diversity is limited, the likelihood surface becomes ambiguous and the parameter estimate may deviate from the true target. In such cases, all FIM-based metrics reflect only the local curvature around the current estimate, causing their trajectory updates to be similarly ineffective.

\begin{figure}[t]
    \centering
    \includegraphics[width=0.9\linewidth]{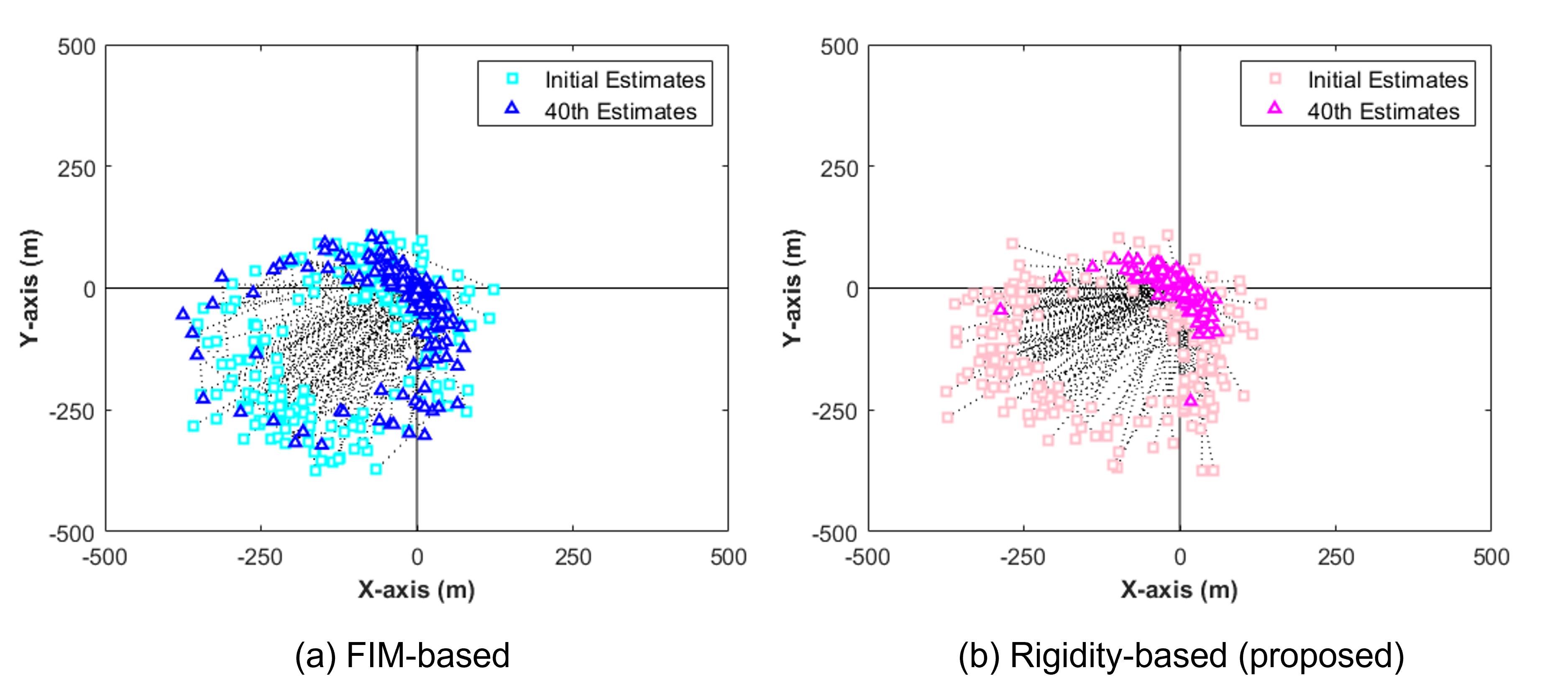}
    \caption{Evolution of target position estimates under (a) the FIM-based method and (b) the proposed rigidity-based method. The proposed method reduces the spatial dispersion of the target estimates more effectively after 40 s, indicating stronger reduction of position ambiguity.}
    \label{fig:ambiguity_reduction}
\end{figure}

To more directly illustrate the ambiguity-reduction behavior of the proposed method, Fig.~\ref{fig:ambiguity_reduction} shows the evolution of target position estimates obtained from Monte Carlo simulations. The square markers denote the initial target estimates at $t=0$ s, while the triangle markers denote the estimates after 40 s. The dotted lines connect the initial and final estimates from the same simulation run. For visualization clarity, 200 Monte Carlo trials are shown. As shown in Fig.~\ref{fig:ambiguity_reduction}(a), the FIM-based method still exhibits a widely dispersed distribution of target estimates after 40 s, indicating that a considerable number of trials remain affected by early-stage position ambiguity. In contrast, Fig.~\ref{fig:ambiguity_reduction}(b) shows that the estimates obtained using the proposed rigidity-based method are more tightly concentrated around the true target position. This result provides empirical evidence that the proposed method reduces position ambiguity through improved geometric conditioning in the considered scenario, rather than a theoretical guarantee that all flip ambiguities or local minima are eliminated.

\begin{figure} 
    \centering
    \includegraphics[width=0.9\linewidth]{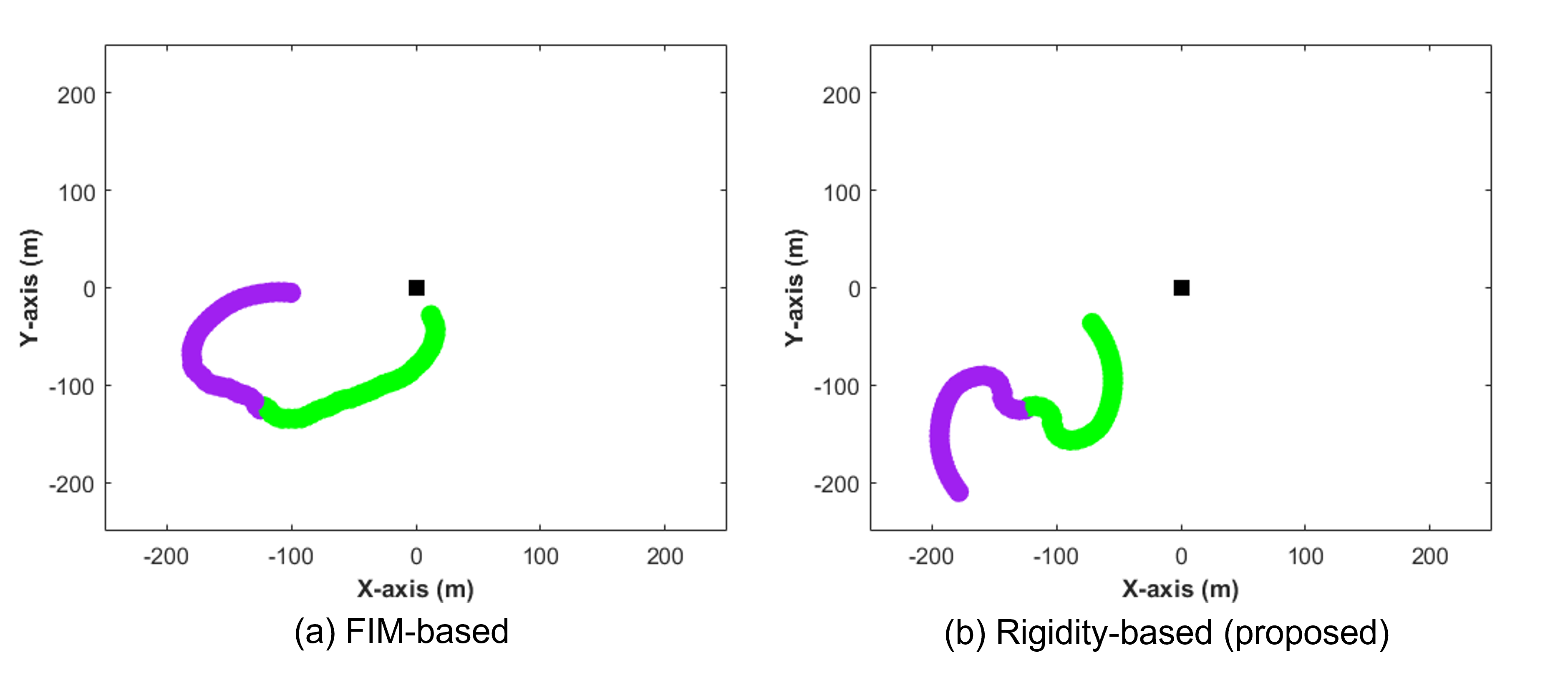}
    \caption{Optimized trajectories of two UAVs in (a) the FIM-based approach using D-optimality and (b) the proposed rigidity-based approach. The purple and green lines represent the UAV paths. The target is located at $(0, 0)$.}
    \label{fig:Trajectories}
\end{figure}

Fig.~\ref{fig:Trajectories} illustrates example UAV trajectories during the first 40 epochs (i.e., 40 seconds) for both the proposed method and the FIM-based method using D-optimality. 
The green and purple lines represent the paths of the two UAVs, while the true target position is indicated by a black square at $(0, 0)$. The rigidity-based method exhibits a distinct behavioral pattern: one UAV tends to move away from the target while the other approaches it, thereby increasing the rigidity of the formation and enhancing localization robustness. 
In contrast, the FIM-based method directs both UAVs toward the estimated target location, aiming to minimize estimation uncertainty according to the FIM criterion. 
However, this strategy can result in less effective spatial exploration and slower convergence in ambiguous scenarios.

\subsection{Scalability with the Number of UAVs}

\begin{figure} 
    \centering
    \includegraphics[width=0.6\linewidth]{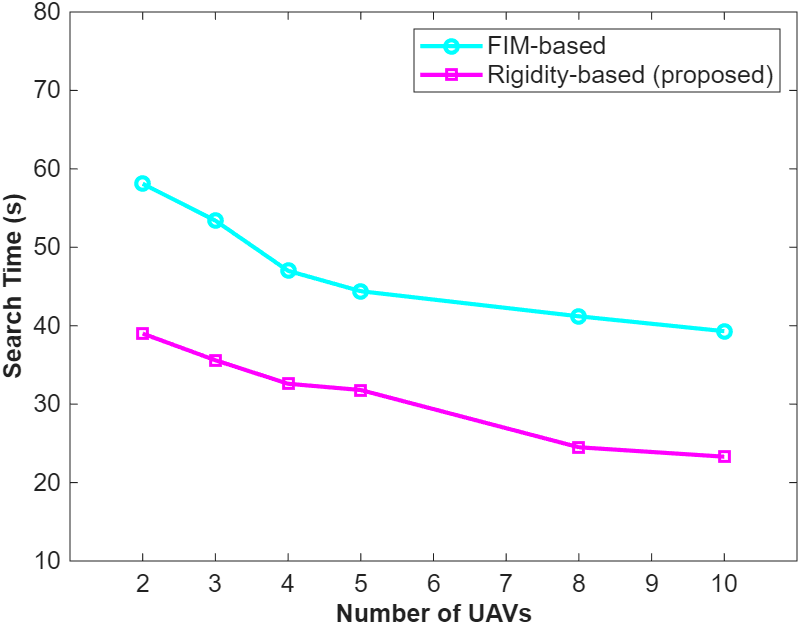}
    \caption{Search time required to achieve an 80\% success rate as the number of UAVs increases from 2 to 10.}
    \label{fig:Scalability}
\end{figure}

This subsection analyzes the scalability of the proposed method with respect to the number of UAVs. 
Fig.~\ref{fig:Scalability} shows the search time required to achieve an 80\% success rate as the number of UAVs increases from 2 to 10. 
As discussed in Section~\ref{subsec:PerformanceComparison}, the choice of optimality criterion has little impact on the performance of FIM-based methods. 
Accordingly, for the remainder of this paper, we report only the results based on the most widely used D-optimality design.

From the results, two key observations can be made. 
First, as the number of UAVs increases, the amount of collected RSS measurements also increases, leading to a general reduction in search time for both methods. 
This indicates that, as expected, a larger UAV swarm contributes to more efficient target localization. Second, the proposed rigidity-based method consistently outperforms the FIM-based method across all UAV counts. 
Even with up to 10 UAVs, the rigidity-based approach achieves shorter search times. 
This trend suggests that the proposed method may continue to exhibit superior performance as the number of UAVs increases.

\subsection{Robustness to UAV Positioning Errors}

This subsection analyzes the robustness of the proposed method to UAV positioning errors. 
In real-world scenarios, such errors inevitably arise when estimating the locations of UAVs, and their magnitude depends on the navigation system employed. To simulate such conditions, we introduced random UAV self-positioning errors following a normal distribution, with representative 95\% positioning-error bounds of 1~m, 3~m, and 20~m. 
The 3~m and 1~m bounds correspond to typical performance levels of global positioning system (GPS)~\cite{DOD20} and differential GPS (DGPS)~\cite{Monteiro05:Accuracy}, respectively, while the 20~m bound represents eLoran-based navigation systems~\cite{Son18:Novel, Kim22:First}. 
This subsection considers the scenario with two UAVs.

\begin{figure} 
    \centering
    \includegraphics[width=0.7\linewidth]{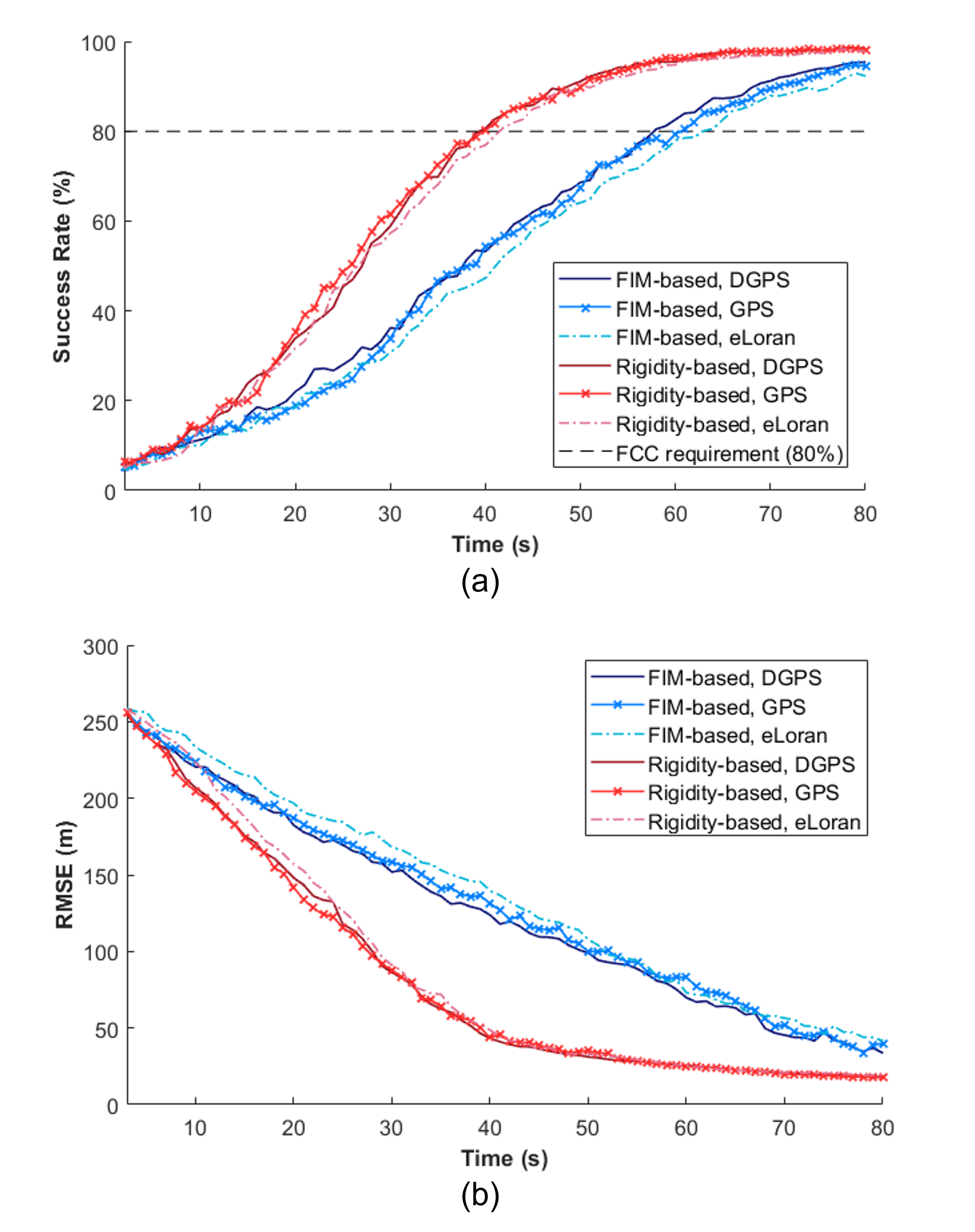}
    \caption{Performance comparison under realistic conditions where UAVs estimate their own positions using DGPS, GPS, and eLoran. 
    (a) Success rate over time, measured as the percentage of simulation iterations that achieved localization within a 50 m error. 
    (b) Localization RMSE over time.}
    \label{fig:Realistic_SuccessRate}
\end{figure}

Fig.~\ref{fig:Realistic_SuccessRate}(a) compares the proportion of simulation runs that achieved a target localization error of less than 50 m over time for both approaches under the GPS, DGPS, and eLoran error scenarios. 
Fig.~\ref{fig:Realistic_SuccessRate}(b) shows the RMSE of the target localization over time.
Table~\ref{tab:Realistic_Time} presents a comparison of the time required to satisfy the FCC E911 requirements (i.e., 50 m accuracy for 80\% of attempts) under realistic positioning conditions. 
In the DGPS, GPS, and eLoran cases, the proposed rigidity-based method reduced the required time by 31.8\%, 34.2\%, and 34.1\%, respectively, compared to the FIM-based approach.

\begin{table} 
\centering \small
\caption{Time Required to Meet FCC E911 Localization Requirements for the Two Approaches Under Realistic Conditions}
\label{tab:Realistic_Time}
\vspace{-4mm}
\begin{center}
{\renewcommand{\arraystretch}{1.4}
 \begin{tabular}[c] {M{2.6cm} | M{1.5cm} | M{1.5cm} | M{1.5cm}}
 \Xhline{1.0pt}
    {} & \thead{DGPS} & \thead{GPS} & \thead{eLoran} \\
    \noalign{\vspace{0pt}}
\hline \vspace{3pt}
 \thead{FIM-based (s)} & \vspace{3pt} 57.8 & \vspace{3pt} 60.6 & \vspace{3pt} 63.0 \\
 \multirow{2}{*}{\thead{Rigidity-based (s)\\(proposed)}} & \multirow{2}{*}{39.4} & \multirow{2}{*}{39.9} & \multirow{2}{*}{41.5} \\
 & & & \\
 \hline
 \thead{Improvement (\%)} & 31.8 & 34.2 & 34.1 \\
 \Xhline{1.0pt}
\end{tabular}}
\end{center}
\end{table}

These results highlight the robustness of the proposed rigidity-based UAV trajectory optimization method to UAV positioning errors. 
Even in scenarios with significant errors of up to 20 m, the proposed method did not exhibit notable performance degradation. This robustness arises from the minor influence of UAV positioning errors on the calculation of the overall rigidity of the framework. 
The proposed rigidity-based approach serves as a method for verifying the uniqueness of a given configuration. 
Even when positioning errors are present, the rigidity of the entire framework can still be assessed using a configuration that accommodates these errors.

\subsection{Sensitivity to Additional NLOS Path Loss}

This subsection presents a sensitivity analysis of the impact of additional NLOS path loss in RSS measurements. 
When obstacles exist between a UAV and the target, additional path loss may occur due to signal blockage. 
Since the severity of this loss can vary depending on the type of obstacle, we investigate how different levels of path loss affect localization performance in terms of search time.
The scenario involves two UAVs. 
It is assumed that an additional NLOS bias ranging from 1 dB to 5 dB is applied to the RSS measurement of one UAV during the time interval between 10 and 20 seconds.

\begin{figure} 
    \centering
    \includegraphics[width=0.6\linewidth]{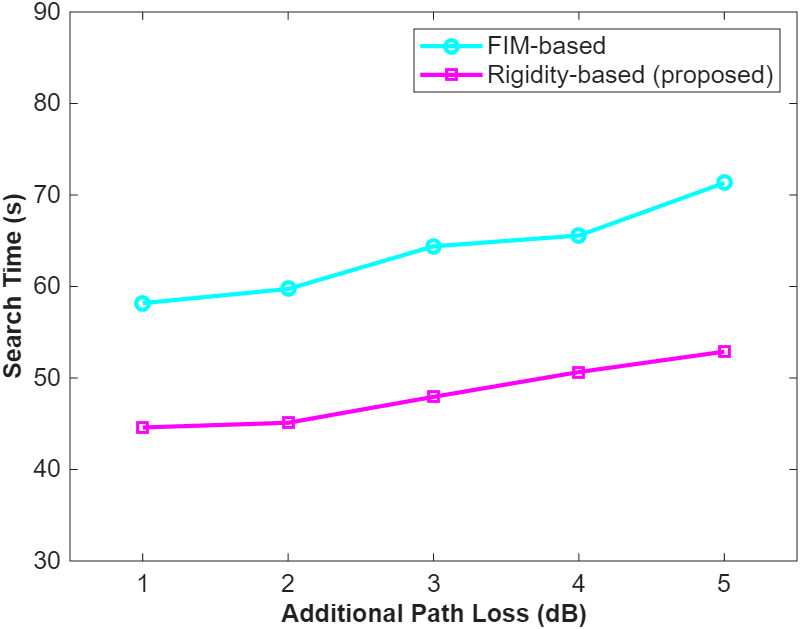}
    \caption{Sensitivity of search time to additional NLOS path loss ranging from 1 dB to 5 dB.}
    \label{fig:Sensitivity}
\end{figure}

Fig.~\ref{fig:Sensitivity} illustrates the resulting search time for each bias level. 
As expected, the search time increases as the magnitude of the NLOS bias increases. 
Nevertheless, across all tested conditions, the proposed rigidity-based method consistently outperforms the FIM-based method in terms of search speed.
It should be noted that the NLOS scenario considered in this study represents a temporary bias affecting a single UAV–target link. In more challenging environments where multiple UAV–target links are simultaneously affected by persistent NLOS conditions, the reliability of RSS measurements may be significantly degraded. In such cases, the performance of RSS-based localization methods---including both FIM-based and rigidity-based trajectory optimization---may deteriorate due to the inherent limitations of the RSS measurement model. Addressing such scenarios would require additional mechanisms such as NLOS detection, bias mitigation, or alternative sensing modalities, which are beyond the scope of this study.

\subsection{Comparison with a PPO-Based Learning Baseline}

To further compare the proposed method with a recent learning-based trajectory planning approach, we implemented a PPO-based baseline~\cite{Schulman2017:Proximal, Raffin2021:Stable}. The PPO policy was trained under the nominal environment with two UAVs, $\sigma_{\rm dB}=5$~dB RSS shadowing, no UAV positioning error, and no additional NLOS bias. The action space was designed to match the candidate heading set used by the FIM-based and rigidity-based methods. Detailed PPO implementation, including the Markov decision process (MDP) formulation, observation space, action space, reward design, network architecture, and hyperparameters, is provided in the supplementary material.

Table~\ref{tab:ppo_comparison} compares the search time required to achieve an 80\% success rate within 50~m. The values in parentheses denote the relative increase compared with the corresponding no-perturbation condition of each method. Under the no-perturbation condition, the PPO-based baseline achieves the shortest search time, indicating that learning-based trajectory planning can be highly effective when the test environment matches the training distribution. However, when evaluated without retraining under degraded conditions, the PPO-based method shows a larger performance degradation. In particular, for NLOS biases of 3~dB or larger and for eLoran-level UAV positioning errors, the proposed rigidity-based method achieves shorter search time and a smaller relative increase. These results suggest that learning-based planning is effective in training-matched scenarios, whereas the proposed rigidity-based method offers stronger robustness under sensing and navigation uncertainties without requiring offline retraining.

It should also be noted that the PPO-based baseline requires offline training before deployment. The PPO policy used for the comparison required 8240.0~s of training, corresponding to approximately 137~min, or 2~h 17~min, on an Apple M4 device with 16~GB memory. In contrast, the proposed rigidity-based method requires no offline training and can be applied directly through online geometric optimization.
To further evaluate learning-based baselines, we additionally trained a domain-randomized PPO policy under randomized NLOS bias and UAV positioning errors, with details and results provided in Appendices~D and~E of the supplementary material. While domain randomization improves the robustness of the PPO baseline, the proposed rigidity-based method shows more graceful performance degradation under severe conditions without requiring offline retraining or prior knowledge of the degradation distribution.

\begin{table}[t]
\centering \small
\caption{Search Time Comparison with the PPO-Based Baseline}
\label{tab:ppo_comparison}
\vspace{-4mm}
\begin{center}
{\renewcommand{\arraystretch}{1.05}
\begin{tabular}[c]{M{2.1cm} | M{2.3cm} | M{2.3cm}}
\Xhline{1.0pt}
    {} & \thead{PPO-based\\(s, increase)} & \thead{Rigidity-based\\(s, increase)} \\
\hline \vspace{3pt}
    \thead{No\\perturbation} & \vspace{8pt} 32.3 (0.0\%) & \vspace{8pt} 39.0 (0.0\%) \\
\hline \vspace{3pt}
    \thead{NLOS 1 dB} & \vspace{3pt} 35.3 (+9.3\%) & \vspace{3pt} 44.6 (+14.4\%) \\
    \thead{NLOS 2 dB} & 42.3 (+31.0\%) & 45.1 (+15.6\%) \\
    \thead{NLOS 3 dB} & 49.3 (+52.6\%) & 47.9 (+22.8\%) \\
    \thead{NLOS 4 dB} & 53.6 (+65.9\%) & 50.7 (+30.0\%) \\
    \thead{NLOS 5 dB} & 57.2 (+77.1\%) & 52.9 (+35.6\%) \\
\hline \vspace{3pt}
    \thead{DGPS} & \vspace{3pt} 36.2 (+12.1\%) & \vspace{3pt} 39.4 (+1.0\%) \\
    \thead{GPS} & 36.5 (+13.0\%) & 39.9 (+2.3\%) \\
    \thead{eLoran} & 55.0 (+70.3\%) & 41.5 (+6.4\%) \\
\Xhline{1.0pt}
\end{tabular}}
\end{center}
\end{table}

\subsection{Sensitivity to Heading Discretization and Turning Constraint}

\begin{figure}[t]
    \centering
    \includegraphics[width=0.55\linewidth]{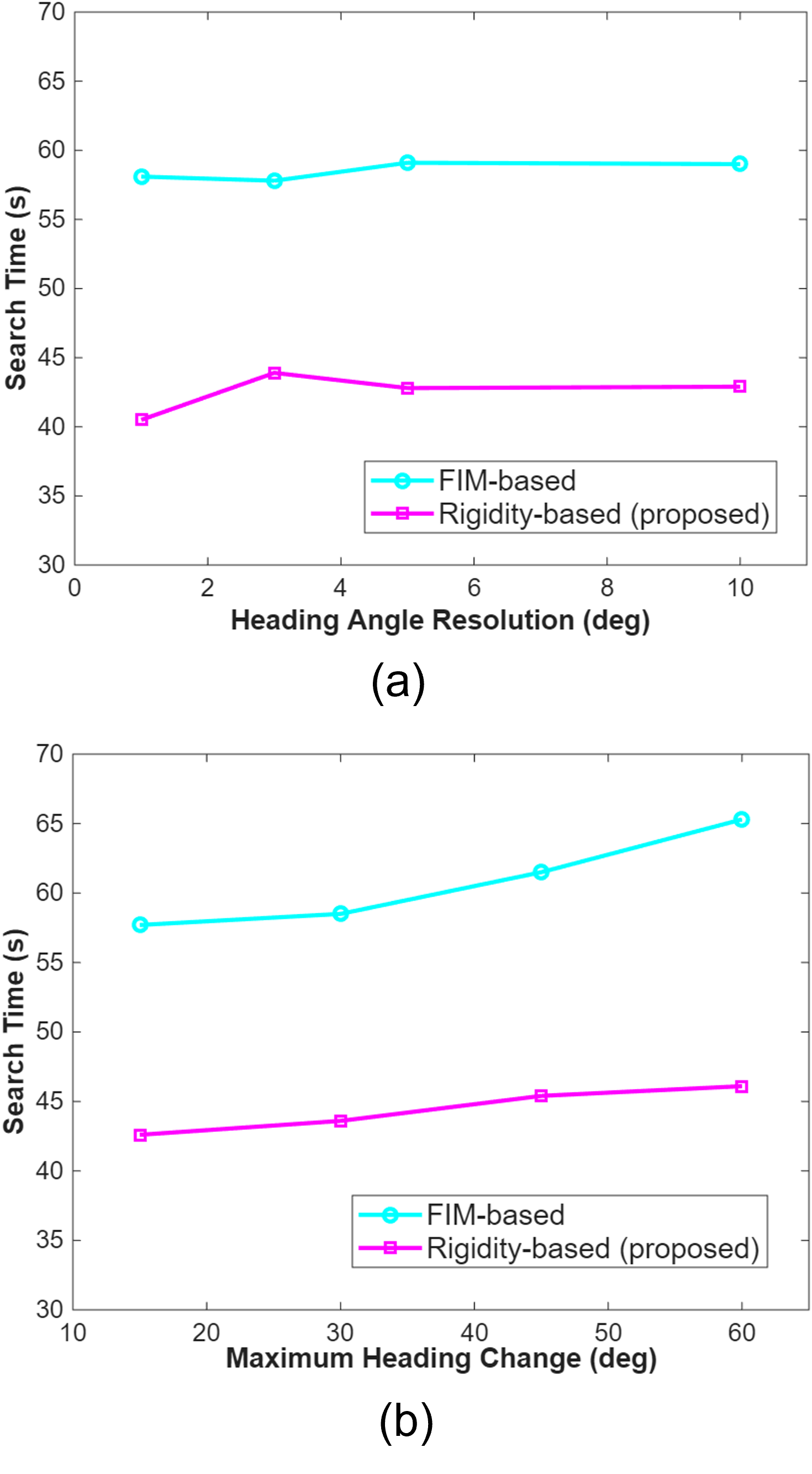}
    \caption{Sensitivity of search time to heading-angle parameters. (a) Search time versus heading angle resolution. (b) Search time versus maximum allowable heading change. The proposed rigidity-based method consistently achieves shorter search time than the FIM-based method across all tested parameter settings.}
    \label{fig:heading_sensitivity}
\end{figure}

The proposed trajectory optimization evaluates a finite set of candidate heading directions and constrains the maximum heading change between consecutive epochs to avoid abrupt UAV maneuvers. To examine the sensitivity to these design choices, we varied the heading angle resolution and the maximum allowable heading change. Fig.~\ref{fig:heading_sensitivity}(a) shows the sensitivity to the heading angle resolution, which was varied as $1^\circ$, $3^\circ$, $5^\circ$, and $10^\circ$. The search time remains relatively stable over this range, and the proposed rigidity-based method consistently outperforms the FIM-based method. Fig.~\ref{fig:heading_sensitivity}(b) shows the sensitivity to the maximum allowable heading change, which was varied as $15^\circ$, $30^\circ$, $45^\circ$, and $60^\circ$. Although the search time increases as the maximum heading change becomes larger in the tested scenario, the proposed method still achieves shorter search time for all tested turning constraints. These results indicate that the performance gain of the proposed method is not strongly dependent on a specific heading discretization or turning-constraint setting.

The increase in search time for larger maximum heading changes reflects a trade-off between maneuvering flexibility and trajectory consistency. Although a larger heading-change limit allows UAVs to rapidly adjust their relative geometry, it can also cause abrupt directional changes when the early-stage MLE fluctuates under noisy RSS measurements. This variability may reduce the stability of the accumulated sensing geometry and require more epochs to obtain sufficiently informative measurements. In contrast, a tighter turning constraint implicitly regularizes the trajectory and promotes smoother geometric improvement. Since the UAV speed is fixed at 5 m/s with a 1 Hz update rate, the trajectory length is proportional to the number of elapsed epochs, so Fig.~\ref{fig:heading_sensitivity} also reflects the trajectory-length trend.

\subsection{Code Execution Time Analysis} \label{subsec:CodeExecutionTimeAnalysis}

\begin{figure} 
    \centering
    \includegraphics[width=0.7\linewidth]{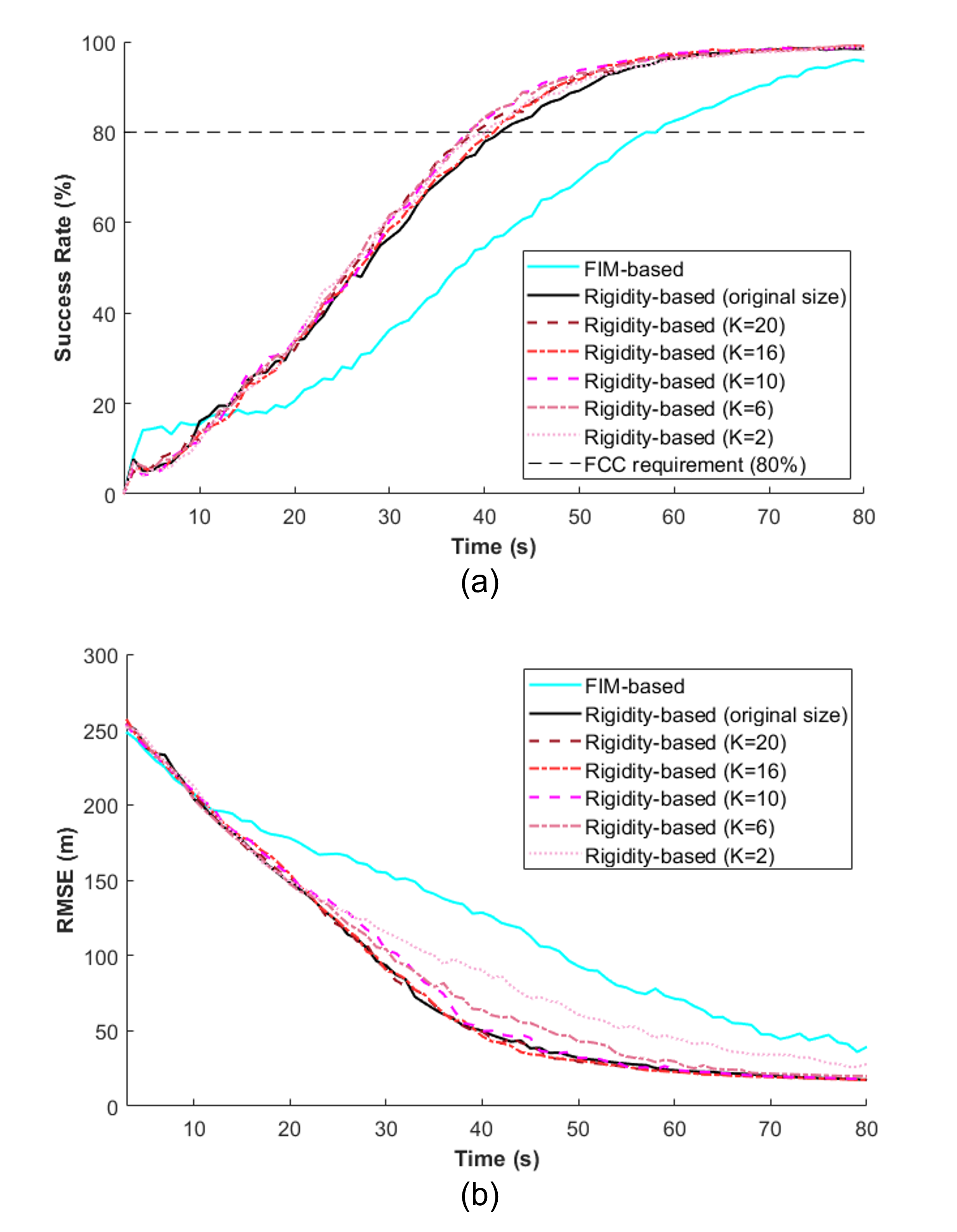}
    \caption{Performance after applying the proposed rigidity matrix size reduction method. 
    (a) Success rate over time, measured as the percentage of simulation iterations that achieved localization within a 50 m error. 
    (b) Localization RMSE over time.}
    \label{fig:SVDRMSE}
\end{figure}

This subsection analyzes the target localization performance and code execution time with respect to varying values of $K$. 
Figs.~\ref{fig:SVDRMSE}(a) and (b) show the time spent to meet the FCC E911 success rate requirement and the target localization RMSE, respectively, for $K = [2, 6, 10, 16, 20]$, compared to using the original size of the rigidity matrix. 
This analysis considers the scenario with two UAVs.
After reducing the size of the rigidity matrix (pruning), we observed a slight increase in success rate but a higher RMSE.
This can be interpreted as follows: pruning the rigidity matrix removed several misleading constraints that had driven some runs to large localization errors, thereby increasing the proportion of runs meeting the 50 m success threshold; however, the accompanying reduction in information slightly inflated errors across many runs, lowering overall precision and increasing the average error.

Misleading constraints are most likely to occur when multiple UAV observations are obtained from geometrically redundant viewpoints, such as near-collinear or clustered configurations. In such situations, additional vertices provide little new geometric diversity and may introduce unstable constraints when measurement noise is present. To quantify the contribution of each vertex, the proposed method evaluates the increase in the smallest nonzero singular value of the rigidity matrix, denoted by $\Delta\sigma_{2|V|-3}$. Since the smallest nonzero singular value reflects the rigidity margin of the sensing framework, $\Delta\sigma_{2|V|-3}$ measures the incremental contribution of each vertex to the geometric conditioning of the localization problem. Measurement noise may affect the target position estimate and consequently the rigidity matrix. However, because the pruning rule evaluates the incremental rigidity contribution of each vertex through $\Delta\sigma_{2|V|-3}$, vertices that provide stronger geometric constraints tend to be retained even in the presence of measurement noise.

\begin{figure} 
    \centering
    \includegraphics[width=0.7\linewidth]{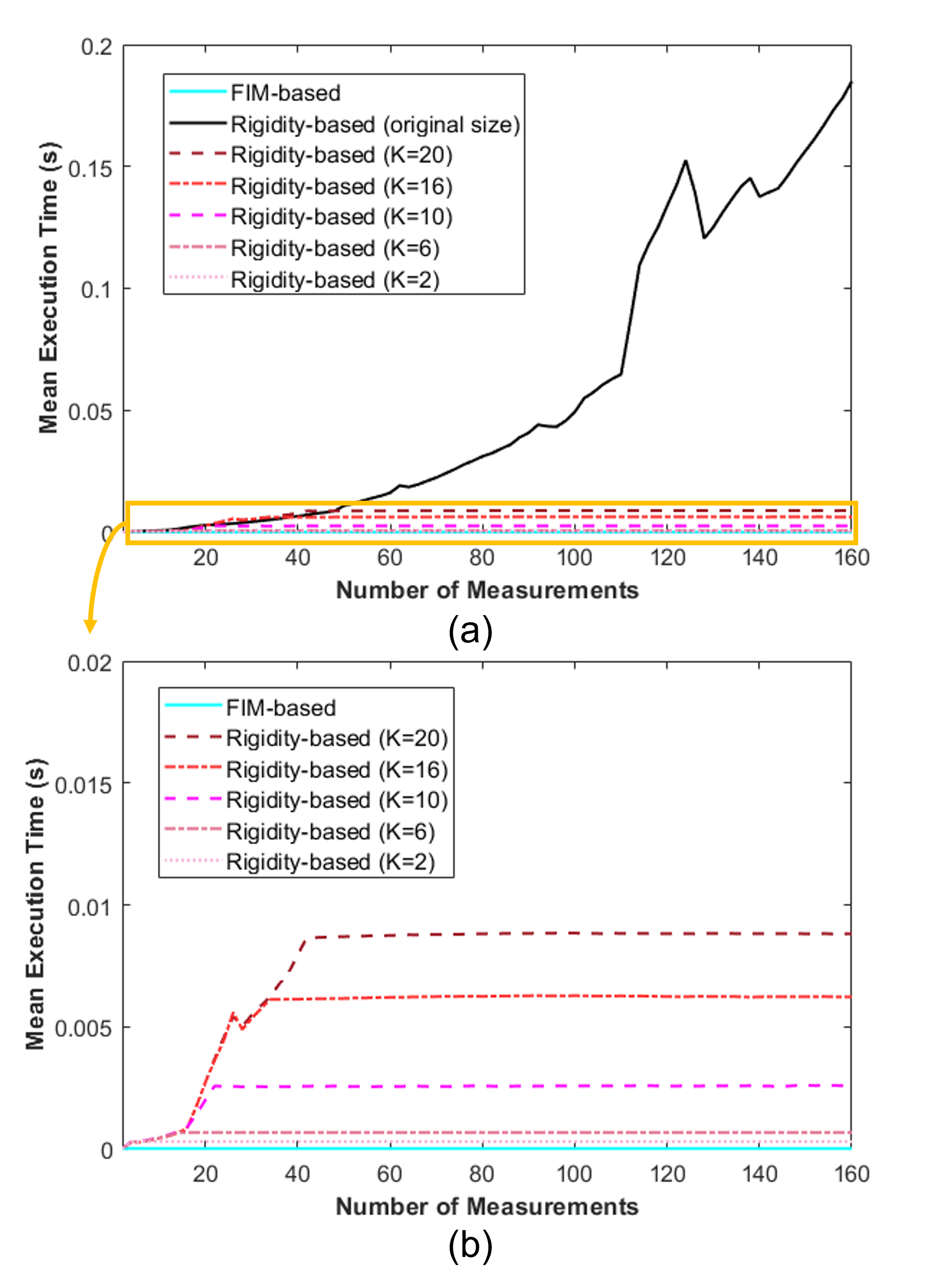}
    \caption{Mean CPU execution time. (b) is an enlarged view of the highlighted region in (a).}
    \label{fig:SVDTime}
\end{figure}

Fig.~\ref{fig:SVDTime} compares the CPU time required to evaluate the cost function in (\ref{eq:ProposedOptimization}). 
The CPU time was measured on a desktop equipped with an Intel Xeon(R) CPU E5-2699 running at 2.30 GHz and 256 GB of memory. 
The reported CPU times were obtained by averaging the results from 1,000 simulation runs. 
All code for the FIM-based and rigidity-based trajectory optimization methods was executed in MATLAB.
Overall, when $K = 20$ and $K = 16$, there is little difference in the RMSE compared to using the original size of the rigidity matrix, while a significant reduction in CPU time is observed. 
However, when $K = 10$, $K = 6$, and $K = 2$, the average RMSEs over time increased by 1.2\%, 1.8\%, and 7.9\%, respectively, compared to that of the original size. 
Despite this, even with $K = 2$, the rigidity-based approach still outperforms the FIM-based approach.
CPU time is greatly reduced through the proposed matrix size reduction. 
CPU times of approximately $2.9 \times 10^{-4}$, $6.6 \times 10^{-4}$, $2.6 \times 10^{-3}$, $6.2 \times 10^{-3}$, and $8.8 \times 10^{-3}$ seconds were maintained when $K = 2$, $K = 6$, $K = 10$, $K = 16$, and $K = 20$, respectively. 
Although the computational time of the rigidity-based approach was higher than that of the simple FIM-based approach, it remained feasible for real-time implementation.

Table~\ref{tab:bigO} summarizes the computational complexity of the different approaches. For the FIM-based method, the objective is based on the determinant of the FIM. Let $d$ denote the dimension of the target state ($d=2$ in this work) and let $N$ denote the number of measurements. Constructing the FIM requires $\mathcal{O}(Nd^2)$ operations, while computing its determinant via matrix factorization requires $\mathcal{O}(d^3)$ operations. Since $d$ is a small constant, the overall complexity scales approximately as $\mathcal{O}(N)$.
In contrast, the rigidity-based method evaluates the singular values of the rigidity matrix $\mathbf{R} \in \mathbb{R}^{m \times n}$. When all past UAV positions are included as vertices, the number of vertices grows over time and the number of edges grows as $\mathcal{O}(|V|^2)$ in the fully connected graph. Computing the full SVD therefore has computational complexity $\mathcal{O}(\max(m,n)\min(m,n)^2)$, which increases as the size of the rigidity matrix grows.

To limit this growth, matrix size reduction can be applied to retain only the $K$ most informative vertices. The size of the rigidity matrix is then bounded as $n = \mathcal{O}(K)$ and $m = \mathcal{O}(K^2)$. Consequently, the computational complexity of the SVD becomes bounded by a constant determined by $K$, meaning that the optimization cost does not increase as measurements accumulate. In addition, the measured execution time required to evaluate the rigidity-based cost function (i.e., a single SVD computation) is on the order of milliseconds, as shown in Fig.~\ref{fig:SVDTime}. Since trajectory optimization evaluates this cost over a finite set of candidate headings once per epoch (1~Hz in our simulations), the overall computational load remains well within real-time processing capability.
While this subsection focuses on the SVD computation time associated with the proposed rigidity-based optimization, a component-wise runtime breakdown of the complete online computational pipeline, covering the MLE grid search, the trajectory optimization stage (candidate heading enumeration and the associated SVD computations), and the heap aggregation, is provided in Appendix~F of the supplementary material.

\begin{table}
\centering \small
\caption{Comparison of Computational Complexity}
\label{tab:bigO}
\vspace{-4mm}
\begin{center}
{\renewcommand{\arraystretch}{1.4}
\begin{tabular}{M{2.9cm} | M{2.3cm} | M{2.3cm}}
\Xhline{1.0pt}

\thead{Method} & \thead{Main Operation} & \thead{Complexity} \\

\hline

\thead{FIM-based}
& \vspace{2pt} FIM construction + determinant
& \vspace{1pt} $\mathcal{O}(N)$ \\

\thead{Rigidity-based\\(original size)}
& \vspace{4pt} SVD
& \vspace{4pt} $\mathcal{O}(mn^2)$ or $\mathcal{O}(nm^2)$ \\

\thead{Rigidity-based\\($K$-vertex reduction)}
& \vspace{4pt} SVD on bounded rigidity matrix
& \vspace{7pt} \shortstack{$\mathcal{O}(K^4)$ \\ (for fixed $K$)} \\

\Xhline{1.0pt}
\end{tabular}}
\end{center}
\end{table}

\subsection{Communication Overhead Analysis}

At each epoch, the $m$-th UAV transmits a compact state packet to the localization server, which can be represented as
\begin{equation}
\mathbf{z}_{m,t} = \left[\mathrm{ID}_m,\; x^\mathrm{UAV}_{m,t},\; y^\mathrm{UAV}_{m,t},\; \hat{P}_{m,t}\right],
\end{equation}
where $\mathrm{ID}_m$ is the UAV identifier, $(x^\mathrm{UAV}_{m,t}, y^\mathrm{UAV}_{m,t})$ denotes the UAV position, and $\hat{P}_{m,t}$ is the RSS measurement at epoch $t$.

Let $b_{\mathrm{ID}}$, $b_{\mathrm{pos}}$, and $b_{\mathrm{RSS}}$ denote the numbers of bits used to represent the identifier, one position coordinate, and one RSS measurement, respectively. Then, the uplink payload transmitted by one UAV per epoch is $B_{\mathrm{UAV}}= b_{\mathrm{ID}} + 2b_{\mathrm{pos}} + b_{\mathrm{RSS}}.$ Accordingly, when $N$ UAVs participate in cooperative localization, the total communication payload per epoch is $B_{\mathrm{tot}}(N) = N\,B_{\mathrm{UAV}},$
which shows that the communication overhead grows linearly with the number of UAVs, i.e., $B_{\mathrm{tot}}(N)=\mathcal{O}(N).$ If the localization update period is $\Delta t$, the required average communication rate is
$R_{\mathrm{comm}}(N)=\frac{B_{\mathrm{tot}}(N)}{\Delta t}
= \frac{N\,B_{\mathrm{UAV}}}{\Delta t}.$

In the proposed setup, $\Delta t = 1$ s and the transmitted information is low-dimensional, consisting only of UAV state variables and scalar RSS measurements. Therefore, for the UAV swarm sizes considered in this work, the required communication load remains modest. Nevertheless, as $N$ increases, delayed or partial information exchange may affect the timeliness of the heading update. In such cases, the rigidity computation may rely on stale or incomplete measurements, which may affect the optimality of the selected heading. 
The proposed framework can still utilize available measurements to maintain geometric diversity for localization. 
This observation is consistent with the results in Section \ref{subsec:CodeExecutionTimeAnalysis}, where the reduced rigidity matrix strategy indicates that removing less informative vertices does not significantly degrade localization performance. 
When the number of UAV agents becomes large and communication or resource constraints become significant, overhead-aware distributed operation strategies, such as those discussed in \cite{Li25:Energy}, may be considered to extend the proposed framework.

\section{Discussion} \label{sec:Discussion}
The present study assumes a stationary target, which is consistent with many emergency scenarios involving injured, trapped, or distressed callers at a fixed location, and is also a common simplification in prior RSS-based UAV localization studies~\cite{Wang18:Performance, Uluskan20:Noncausal, Koohifar16:Receding}. Extending the proposed framework to moving or intermittently relocating targets is an important direction for broader deployment. For continuously moving targets, the state can be augmented to position and velocity, $\mathbf{x}_{{\rm tar},t}=[x_t,y_t,\dot{x}_t,\dot{y}_t]^T$, and recursively updated using tracking filters (e.g., extended Kalman filter, particle filter, or moving-horizon estimator). The rigidity-based trajectory optimization can then be performed in a receding-horizon manner using the predicted target position. 
For intermittently relocating targets, residual-based change detection or multiple-hypothesis tracking can be combined with the proposed framework to identify abrupt relocations and reinitialize the trajectory planner around the updated target hypothesis. These extensions involve additional motion modeling and theoretical analysis of time-varying frameworks, which are left as future work.

The current formulation does not explicitly consider UAV--UAV collision avoidance or minimum safety-distance constraints. In practical multi-UAV deployments, these constraints can be incorporated by replacing the original candidate heading set in Eq.~\eqref{eq:ProposedOptimization} with a safety-aware feasible set, $\mathcal{A}_{m,t}^{\rm safe}$. Here, $d_{\rm safe}$ denotes the prescribed minimum separation distance, and any candidate heading that leads to $\|\mathbf{x}_{m,t+1}^{\rm UAV}-\mathbf{x}_{j,t+1}^{\rm UAV}\|<d_{\rm safe}$ for another UAV $j$ is excluded from $\mathcal{A}_{m,t}^{\rm safe}$. Alternatively, collision avoidance can be included as a penalty term in the trajectory optimization objective. A full treatment of safety-aware rigidity-based planning, including dynamic collision avoidance and feasibility analysis, is left for future work.

While this study provides comprehensive numerical validation, future work will include hardware-in-the-loop (HIL) simulations and field experiments to further evaluate the proposed framework under realistic communication delays, sensing uncertainties, and hardware constraints.

\section{Conclusion}
\label{sec:Conclusion}

Accurate and timely localization of emergency callers is critical for enabling effective response within vehicular technology and mission-critical wireless systems.
To address this challenge, this paper optimized the trajectories of multiple UAVs for time-constrained localization tasks. Traditional FIM-based UAV trajectory optimization approaches become less effective in scenarios prone to target-position ambiguity, such as real-world rescue operations in which UAVs typically begin their missions from a single base and lack geometric diversity.

Motivated by the concept of rigidity, we developed a rigidity-based UAV trajectory optimization method that rapidly reduces position ambiguity and shortens the time required to meet emergency localization requirements. Our findings indicate that rigidity serves as a reliable optimization metric when ambiguity is present, guiding UAVs toward formations that promote unique and informative geometric configurations.

Simulation results demonstrated a 32.9\% reduction in search time over existing FIM-based strategies for meeting FCC horizontal emergency localization requirements. The method also scales effectively to UAV swarms of up to ten agents and shows robustness to UAV positioning errors and moderate NLOS path-loss conditions in the considered simulation scenarios. Additional sensitivity analysis showed that the performance gain is not strongly dependent on a specific heading discretization or turning-constraint setting. Compared with a PPO-based learning baseline, the proposed training-free method demonstrated stronger robustness under increased NLOS biases and UAV positioning errors without requiring offline retraining. Furthermore, computational-time analysis confirmed its feasibility for real-time implementation.

\section*{Acknowledgment}

Generative AI was used solely to assist with grammar and language improvements during the manuscript preparation process.  
No content, ideas, data, or citations were generated by AI.  
All technical content, methodology, analysis, and conclusions were written and verified solely by the authors.

\bibliographystyle{IEEEtran}
\bibliography{mybibfile, IUS_publications}

\begin{IEEEbiography}[{\includegraphics[width=1.3in,height=1.3in,clip,keepaspectratio]{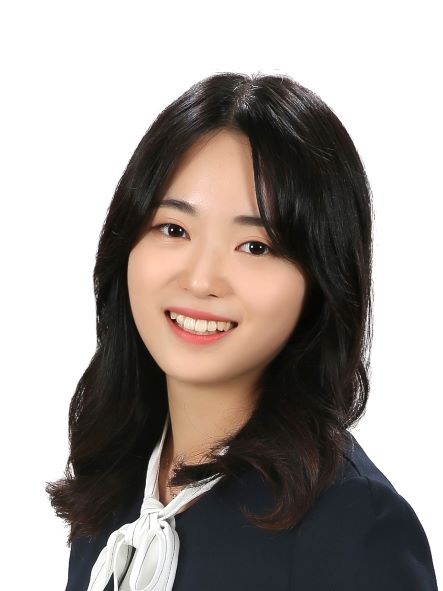}}]{Halim Lee} is a Ph.D. candidate in the School of Integrated Technology, Yonsei University, Incheon, Republic of Korea. She received the B.S. degree in Integrated Technology from Yonsei University. Her research interests include target localization and tracking, alternative PNT systems, and intelligent unmanned systems.
Ms. Lee received the Undergraduate and Graduate Fellowships from the Information and Communications Technology (ICT) Consilience Creative Program supported by the Ministry of Science and ICT, Republic of Korea. 
\end{IEEEbiography}

\begin{IEEEbiography}[{\includegraphics[width=1in,height=1.25in,clip,keepaspectratio]{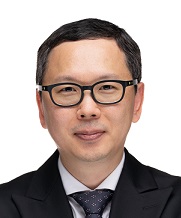}}]{Jiwon Seo} (Senior Member, IEEE) received a B.S. degree in mechanical engineering (division of aerospace engineering) from the Korea Advanced Institute of Science and Technology (KAIST), Daejeon, Republic of Korea, in 2002.
He earned M.S. degrees in aeronautics and astronautics (2004) and electrical engineering (2008), as well as a Ph.D. in aeronautics and astronautics (2010), all from Stanford University, Stanford, CA, USA.
He is currently an Underwood Distinguished Professor at Yonsei University, where he is a professor in the School of Integrated Technology, Incheon, Republic of Korea. 
His research interests include GNSS anti-jamming technologies, complementary PNT systems, and intelligent unmanned systems.
Dr. Seo is a member of the International Advisory Council of the Resilient Navigation and Timing Foundation, Alexandria, VA, USA, and the Advisory Committee on Defense of the Presidential Advisory Council on Science and Technology, Republic of Korea.
\end{IEEEbiography}

\vfill

\end{document}


\title{\textit{Supplementary Material for}\\
Rigidity-Based Multi-UAV Trajectory Optimization for Rapid Cooperative Emergency Target Localization}

\author{Halim~Lee and Jiwon~Seo,~\IEEEmembership{Senior Member,~IEEE}}

\maketitle

\begin{strip}
\noindent\textbf{Description:} This document provides supplementary details for the manuscript ``Rigidity-Based Multi-UAV Trajectory Optimization for Rapid Cooperative Emergency Target Localization.'' It describes the PPO-based learning baseline used in Section~VII-E, covering both the nominally trained and domain-randomized variants, and reports the robustness evaluation of the domain-randomized baseline. It also presents a component-wise runtime analysis of the complete closed-loop computational pipeline, complementing the discussion in Section~VII-G.
 
\vspace{1ex}
\noindent\rule{\textwidth}{0.4pt}
\vspace{0.5ex}
\end{strip}

\appendices

\section{Problem Formulation as a Markov Decision Process} \label{app:MDP}

The cooperative UAV target localization problem is formulated as an MDP $\mathcal{M}=(\mathcal{S},\mathcal{A},P,R,\gamma)$, where $\mathcal{S}$ denotes the state space, $\mathcal{A}$ the action space, $P$ the transition dynamics induced by the UAV motion and the RSS measurement model, $R$ the reward function, and $\gamma\in[0,1)$ the discount factor. The agent observes a partial state through a structured observation vector and selects discrete heading angles for all UAVs at each epoch.

\subsection{Observation Space}

At each epoch $t$, the observation vector $\mathbf{o}_t\in\mathbb{R}^{d_o}$ consists of four components:
\begin{itemize}
    \item Current UAV positions normalized by the map size: $\bar{\mathbf{x}}^{\rm UAV}_{m,t}=\mathbf{x}^{\rm UAV}_{m,t}/L_{\rm map}$, for $m=1,\dots,M$, where $L_{\rm map}=375$~m.
    \item Recent RSS history of length $N_{\rm hist}=5$ epochs across all UAVs, normalized as $(\hat{P}_{m,t}+25)/30$ to roughly map the expected RSS range to $[-1,1]$.
    \item Current target estimate $\hat{\mathbf{x}}_{\rm tar}/L_{\rm map}$.
    \item Normalized time progress $t/T_{\max}$, where $T_{\max}=100$ epochs.
\end{itemize}
The total observation dimension for $M=2$ UAVs is $d_o=2M+N_{\rm hist}M+2+1=17$.

\subsection{Action Space}

To match the discretization of the proposed rigidity-based method, the action space is defined as a multi-discrete heading-offset selection. At each epoch, the agent selects an action $a_{m,t}\in\{0,1,\dots,N_a-1\}$ for each UAV $m$, which maps to a heading offset $\Delta\alpha_{m,t}\in\{-\alpha_{\rm lim},\,-\alpha_{\rm lim}+\delta,\,\dots,\,\alpha_{\rm lim}\}$ relative to the previous heading. Here $\alpha_{\rm lim}=20^\circ$ is the maximum allowable heading change per epoch, and $\delta=1^\circ$ is the heading discretization step, yielding $N_a=41$ candidate actions per UAV.
These values match the candidate-heading set used by the rigidity-based and FIM-based methods evaluated in the main manuscript, ensuring a fair comparison. The joint action space is thus $\mathcal{A}=\{0,\dots,40\}^M$.

\subsection{Reward Function}

The reward function combines a dense localization-error term and a sparse bonus tied to the FCC criterion:
\begin{equation}
    r_t = -\frac{\|\hat{\mathbf{x}}_{{\rm tar},t}-\mathbf{x}_{\rm tar}\|}{100}
    + \mathbb{I}\left[\|\hat{\mathbf{x}}_{{\rm tar},t}-\mathbf{x}_{\rm tar}\|<50\right]\cdot 10,
\end{equation}
where $\hat{\mathbf{x}}_{{\rm tar},t}$ is the target estimate obtained via maximum likelihood estimation (using the same grid-search MLE as the proposed method), $\mathbf{x}_{\rm tar}$ is the true target position, and $\mathbb{I}[\cdot]$ is the indicator function. The dense term provides a shaped learning signal proportional to localization accuracy, while the sparse bonus directly encourages meeting the FCC horizontal positioning requirement. It should be noted that the true target position is used only by the simulator to compute the reward during offline training. During evaluation, the trained policy receives only the observation vector defined above and does not require access to the true target position.

\subsection{Episode Definition}

Each training episode starts with both UAVs at the base location $(-125,-125)$~m and runs for $T_{\max}=100$ epochs, corresponding to 100~s at a 1~Hz update rate. The target is fixed at the origin during training, matching the in-distribution evaluation scenario described in Section~VII-E of the main manuscript.

\section{PPO Algorithm, Hyperparameters, and Training Time} \label{app:ppo}

We adopt the PPO implementation from the Stable-Baselines3 library~\cite{stable_baselines3} with the network architecture and hyperparameters summarized in Table~\ref{tab:ppo_hyperparams}. The policy network $\pi_\theta(\mathbf{a}|\mathbf{o})$ and value network $V_\phi(\mathbf{o})$ are implemented as separate multi-layer perceptrons (MLPs) with two hidden layers of 128 units each and Tanh activations. For the multi-discrete action space, the policy outputs independent categorical distributions over heading offsets for each UAV. PPO training was performed using random seed 0, and the trained policy was evaluated over 1,000 Monte Carlo simulations per scenario.

\begin{table}[htbp]
\centering
\caption{PPO Hyperparameters}
\label{tab:ppo_hyperparams}
\begin{tabular}{ll}
\toprule
\textbf{Parameter} & \textbf{Value} \\
\midrule
Learning rate & $3\times 10^{-4}$ \\
Number of parallel environments & 8 \\
Rollout length per environment ($n_{\rm steps}$) & 2048 \\
Mini-batch size & 64 \\
Number of epochs per update & 10 \\
Discount factor ($\gamma$) & 0.99 \\
GAE parameter ($\lambda$) & 0.95 \\
Clipping range ($\epsilon$) & 0.2 \\
Entropy coefficient & 0.01 \\
Value-function coefficient & 0.5 \\
Maximum gradient norm & 0.5 \\
Policy network architecture & MLP, [128, 128], Tanh \\
Value network architecture & MLP, [128, 128], Tanh \\
Total training timesteps & $1\times 10^{6}$ \\
Random seed & 0 \\
\bottomrule
\multicolumn{2}{l}{\footnotesize GAE: generalized advantage estimation;}\\
\multicolumn{2}{l}{\footnotesize MLP: multi-layer perceptron.}
\end{tabular}
\end{table}

Training was performed on an Apple M4 device with 16~GB memory. The PPO policy used for the comparison required 8240.0~s of training, corresponding to approximately 137~min, or 2~h 17~min. By contrast, the proposed rigidity-based method requires \emph{no offline training} and can be applied directly through online geometric optimization. In addition, the online computation time of the proposed method remains feasible for real-time operation, as discussed in Section~VII-G of the main manuscript. This is a key practical distinction: the PPO baseline may require retraining when the deployment environment or task specification changes, whereas the rigidity-based method is immediately deployable without offline policy training.

\section{Evaluation Setting Specifications}

\subsection{In-Distribution Evaluation}

The PPO policy was trained under the nominal environment with two UAVs, $\sigma_{\rm dB}=5$~dB RSS shadowing, no UAV positioning error, no additional NLOS bias, and the target located at the origin. The in-distribution evaluation in Section~VII-E of the main manuscript uses the same environment parameters: both UAVs start from $(-125,-125)$~m, the target is located at the origin, and no UAV positioning errors or additional NLOS bias are applied. This setting is the most favorable scenario for PPO since the test distribution matches the training distribution.

\subsection{Robustness Evaluation}

For the robustness evaluation, the PPO policy trained in the nominal environment was evaluated under the following perturbed conditions without retraining or fine-tuning, as described in Section~VII-E of the main manuscript:
\begin{itemize}
    \item \textit{UAV positioning errors}: Gaussian UAV self-positioning errors are added to represent DGPS-, GPS-, and eLoran-level positioning uncertainties. The error magnitudes are chosen to correspond approximately to 95\% positioning-error bounds of 1~m, 3~m, and 20~m, respectively.
    \item \textit{Additional NLOS path loss}: A bias of 1--5~dB is applied to the RSS measurement of one UAV during the time interval between 10~s and 20~s, matching the protocol in Section~VII-D of the main manuscript.
\end{itemize}

\section{Domain-Randomized PPO Baseline}
\label{app:dr_ppo}

In the main manuscript, the PPO baseline is trained under the nominal environment and evaluated under degraded conditions without retraining. As noted in Section~VII-E, this nominal-trained PPO reflects a realistic deployment scenario in which the perturbation conditions are not known in advance. To provide a complementary learning-based comparison, this appendix describes a domain-randomized PPO (PPO-DR) baseline, in which the perturbation parameters are randomized during training so that the policy is explicitly exposed to degraded sensing and navigation conditions. The corresponding evaluation results are reported in Appendix~\ref{app:dr_ppo_result}.

The PPO-DR baseline uses the same MDP formulation, observation space, action space, reward function, network architecture, and hyperparameters as the nominal PPO baseline described in Appendices~\ref{app:MDP} and~\ref{app:ppo}. The only difference is that the UAV positioning error and the additional NLOS path loss are re-sampled at the beginning of every training episode, rather than being fixed to the nominal (error-free) condition. In this way, the policy is trained over a distribution of degraded conditions instead of a single nominal environment.

\subsection{Domain Randomization Scheme}

At the beginning of each training episode, the perturbation parameters are sampled as follows.

\subsubsection{UAV Positioning Error}
The standard deviation of the UAV self-positioning error is sampled from a uniform distribution,
\begin{equation}
    \sigma_{\mathrm{pos}} \sim \mathcal{U}(0, \sigma_{\mathrm{pos}}^{\max}),
\end{equation}
where $\sigma_{\mathrm{pos}}^{\max}$ is set to cover the most severe positioning condition considered in the main manuscript (eLoran-level errors). During the episode, a zero-mean Gaussian perturbation with the sampled standard deviation $\sigma_{\mathrm{pos}}$ is added independently to the reported position of each UAV at every epoch. Consistent with the convention in Section~VII-C of the main manuscript, the true UAV positions are used for the physical RSS measurements, while the perturbed (reported) positions are used for target estimation and in the observation vector.

\subsubsection{Additional NLOS Path Loss}
With probability $p_{\mathrm{NLOS}}$, an additional NLOS path loss is activated for the episode; otherwise, no NLOS bias is applied. When NLOS is active, the bias magnitude is sampled from a uniform distribution,
\begin{equation}
    b_{\mathrm{NLOS}} \sim \mathcal{U}(0, b_{\mathrm{NLOS}}^{\max}),
\end{equation}
and is applied to the RSS measurement of a randomly selected UAV over a randomly placed time window within the episode. This randomized protocol generalizes the fixed single-link, single-window NLOS scenario used in Section~VII-D of the main manuscript, exposing the policy to a range of NLOS magnitudes, affected links, and onset times.

\begin{table}[!t]
    \centering
    \caption{Domain Randomization Parameters for the PPO-DR Baseline}
    \label{tab:dr_params}
    \renewcommand{\arraystretch}{1.3}
    \begin{tabular}{lc}
        \hline
        \textbf{Parameter} & \textbf{Value} \\
        \hline        $\sigma_{\mathrm{pos}}^{\max}$ & $10$ m \\
        $b_{\mathrm{NLOS}}^{\max}$ & $5$ dB \\
        $p_{\mathrm{NLOS}}$ & $0.5$ \\
        NLOS window length & randomly placed within episode \\
        NLOS-affected UAV & randomly selected per episode \\
        \hline
    \end{tabular}
\end{table}

Table~\ref{tab:dr_params} summarizes the domain randomization parameters. The ranges are chosen to span the degraded conditions evaluated in the main manuscript, so that the perturbation magnitudes encountered at test time lie within the training distribution. All other environment parameters, including the number of UAVs, RSS shadowing standard deviation, candidate heading set, and episode length, are identical to those of the nominal PPO baseline.

\subsection{Training and Evaluation Protocol}

The PPO-DR policy is trained with the same total number of timesteps, optimizer settings, and network architecture as the nominal PPO baseline, using random seed
0. The only modification is the per-episode domain randomization described above. The trained PPO-DR policy is then evaluated under the same fixed conditions and Monte Carlo protocol used for the nominal PPO baseline in Section~VII-E, and the resulting comparison is presented in Appendix~\ref{app:dr_ppo_result}.

\section{Evaluation Results of the Domain-Randomized PPO Baseline}
\label{app:dr_ppo_result}

This appendix reports the robustness evaluation of the PPO-DR baseline described in Appendix~\ref{app:dr_ppo}. For reference, the results are compared against both the nominally trained PPO baseline (PPO-nominal) reported in Section~VII-E of the main manuscript and the proposed rigidity-based method. All methods are evaluated under identical test conditions, namely the no-perturbation condition, the additional NLOS path-loss conditions from $1$ dB to $5$ dB, and the DGPS-, GPS-, and eLoran-level UAV positioning-error conditions. For each method and condition, the search time required to reach an $80\%$ success rate within $50$ m is computed over $1{,}000$ Monte Carlo simulations. The values in parentheses denote the relative increase in search time with respect to the no-perturbation condition of each method.

\begin{table}[!t]
    \centering
    \caption{Search Time Comparison Including the Domain-Randomized PPO Baseline}
    \label{tab:dr_ppo_comparison}
    \renewcommand{\arraystretch}{1.2}
    \begin{tabular}{lccc}
        \hline
        \textbf{Condition} & \textbf{PPO-nominal} & \textbf{PPO-DR} &
        \textbf{Rigidity-based} \\
        & \textbf{(s, increase)} & \textbf{(s, increase)} & \textbf{(s, increase)} \\
        \hline
        \textbf{No perturbation} & 32.3 (0.0\%)  & 31.6 (0.0\%)  & 39.0 (0.0\%)  \\
        \hline
        \textbf{NLOS 1 dB} & 35.3 (+9.3\%)  & 33.9 (+7.3\%)  & 44.6 (+14.4\%) \\
        \textbf{NLOS 2 dB} & 42.3 (+31.0\%) & 36.0 (+13.9\%) & 45.1 (+15.6\%) \\
        \textbf{NLOS 3 dB} & 49.3 (+52.6\%) & 37.6 (+19.0\%) & 47.9 (+22.8\%) \\
        \textbf{NLOS 4 dB} & 53.6 (+65.9\%) & 41.4 (+31.0\%) & 50.7 (+30.0\%) \\
        \textbf{NLOS 5 dB} & 57.2 (+77.1\%) & 44.7 (+41.5\%) & 52.9 (+35.6\%) \\
        \hline
        \textbf{DGPS}   & 36.2 (+12.1\%) & 32.4 (+2.5\%)  & 39.4 (+1.0\%) \\
        \textbf{GPS}    & 36.5 (+13.0\%) & 32.7 (+3.5\%)  & 39.9 (+2.3\%) \\
        \textbf{eLoran} & 55.0 (+70.3\%) & 41.4 (+31.0\%) & 41.5 (+6.4\%) \\
        \hline
    \end{tabular}
\end{table}

Table~\ref{tab:dr_ppo_comparison} summarizes the results. Two observations can be made.
First, domain randomization clearly improves the robustness of the learning-based baseline relative to nominal training. Compared with PPO-nominal, the PPO-DR baseline achieves smaller relative increases in search time under almost all degraded conditions. For example, under the $5$ dB NLOS bias, the relative increase decreases from $+77.1\%$ for PPO-nominal to $+41.5\%$ for PPO-DR, and under the eLoran-level positioning error it decreases from $+70.3\%$ to $+31.0\%$.

Second, even with domain randomization, the proposed rigidity-based method exhibits more stable degradation behavior than PPO-DR as the conditions worsen. Although PPO-DR attains shorter absolute search times, its relative degradation remains larger than that of the rigidity-based method in the more severe cases. For the $5$ dB NLOS bias, the relative increase is $+41.5\%$ for PPO-DR versus $+35.6\%$ for the rigidity-based method, and for the eLoran-level positioning error it is $+31.0\%$ for PPO-DR versus only $+6.4\%$ for the rigidity-based method. The rigidity-based method therefore maintains a smaller relative performance loss under the most challenging sensing and navigation conditions.

These results indicate that domain randomization is an effective way to improve the robustness of learning-based trajectory planning when the degradation distribution is known and can be reproduced during training. However, this requires prior knowledge of the degradation characteristics and an offline training stage, and the policy may need to be retrained when the deployment conditions change. In contrast, the proposed rigidity-based method achieves comparable and, in terms of relative degradation, more stable robustness without any offline training or prior specification of the degradation distribution. This is a key practical advantage for emergency localization, where the propagation conditions, NLOS severity, and UAV positioning-error levels are typically unknown in advance.

\section{Total Pipeline Runtime Analysis}
\label{app:runtime}

To complement the SVD timing analysis in Section~VII-G of the main manuscript, this appendix reports the runtime of the complete computational pipeline of the proposed closed-loop online operation. At each epoch, the pipeline consists of three computational stages: (i) grid-search MLE of the target position, (ii) rigidity-based trajectory optimization over the candidate headings of all UAVs, and (iii) min-heap aggregation for the rigidity matrix size reduction described in Section~V-D of the main manuscript.

\begin{figure}[!t]
    \centering
    \includegraphics[width=0.9\linewidth]{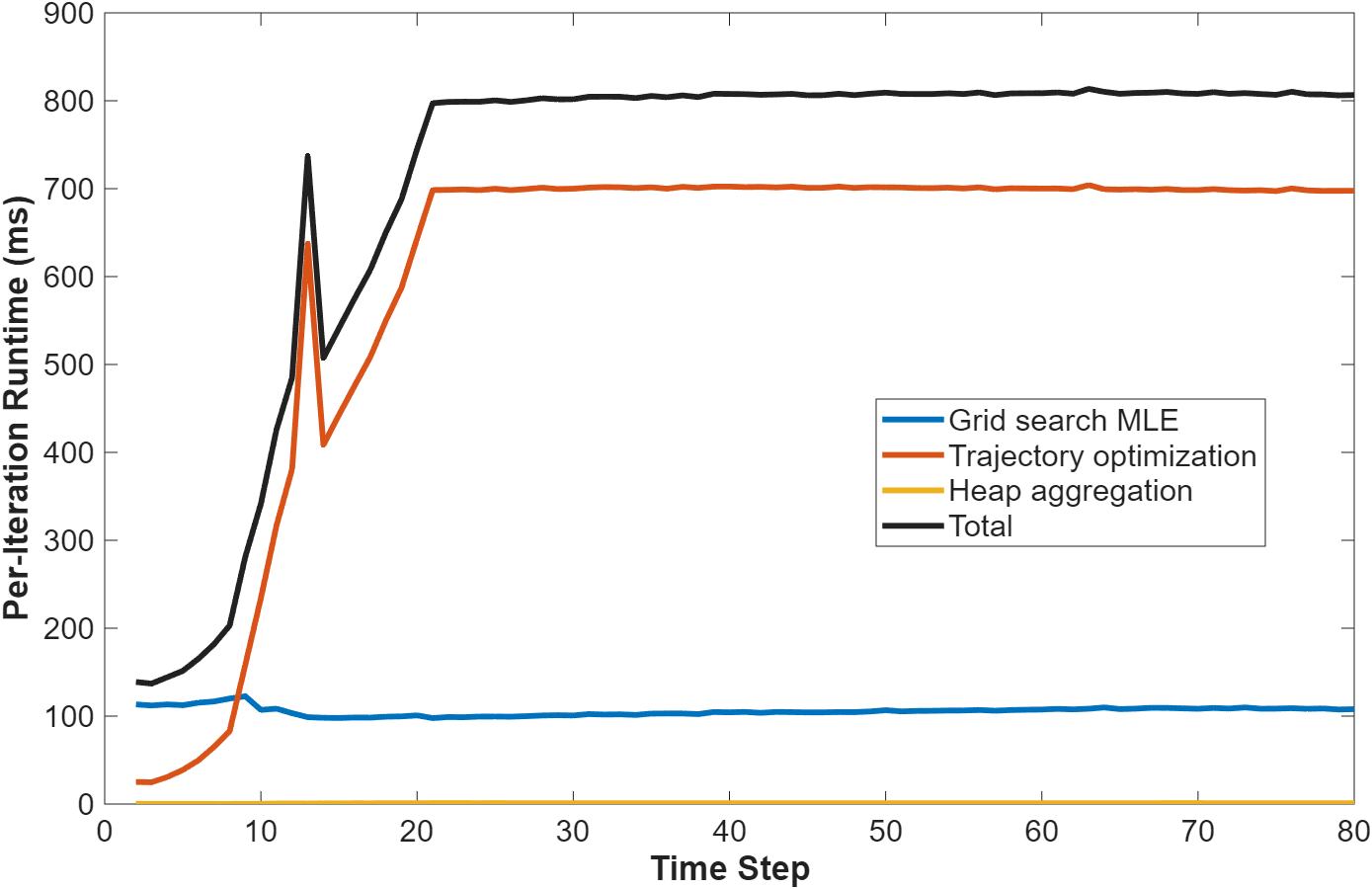}
    \caption{Per-epoch runtime of the complete computational pipeline, broken down into the grid-search MLE, rigidity-based trajectory optimization, and heap aggregation stages. The per-epoch runtime grows during the early epochs and saturates after approximately $20$~s once the rigidity matrix size reduction reaches the desired number of vertices.}
    \label{fig:runtime_curves}
\end{figure}

The runtime of each stage was measured separately over $100$ independent runs. For each stage, we report the mean, standard deviation, $95$th percentile, and $99$th percentile of the per-epoch runtime. Because the size of the rigidity matrix grows during the early epochs and then saturates once the matrix size reduction reaches the desired number of vertices $K=20$, the per-epoch runtime stabilizes after approximately $20$~s, as shown in Fig.~\ref{fig:runtime_curves}. Accordingly, the statistics reported in Table~\ref{tab:runtime} are computed over the saturated regime ($t > 20$~s), which represents the steady-state computational load of the pipeline.

All measurements were obtained on a desktop equipped with an Intel Xeon(R) CPU E5-2699 running at $2.30$~GHz with $256$~GB of memory, identical to the platform used in Section~VII-G of the main manuscript. The grid-search MLE, consistent with the assumption adopted throughout this paper, was parallelized using the MATLAB \texttt{parfor} construct with $36$ workers. The rigidity-based trajectory optimization stage evaluates the smallest nonzero singular value of the rigidity matrix for every candidate heading of each UAV; in the current implementation, the UAVs are processed sequentially, so this stage can be further accelerated by parallelizing the per-UAV optimization. The heap aggregation stage corresponds to the time required to maintain the min-heap used for the rigidity matrix size reduction.

\begin{table}[!t]
    \centering
    \caption{Per-Epoch Runtime Breakdown of the Complete Computational Pipeline in the Saturated Regime ($t > 20$~s)}
    \label{tab:runtime}
    \renewcommand{\arraystretch}{1.3}
    \begin{tabular}{lcccc}
        \hline
        \textbf{Stage} & \textbf{Mean} & \textbf{Std} & \textbf{95th} & \textbf{99th} \\
        & \textbf{(ms)} & \textbf{(ms)} & \textbf{(ms)} & \textbf{(ms)} \\
        \hline
        Grid search MLE          & 105.16 & 6.90 & 117.96 & 123.42 \\
        Trajectory optimization  & 700.14 & 8.69 & 712.34 & 717.28 \\
        Heap aggregation         & 1.00   & 0.10 & 1.05   & 1.10   \\
        \hline
        \textbf{Total}           & \textbf{806.29} & \textbf{12.32} & \textbf{826.75} & \textbf{834.98} \\
        \hline
    \end{tabular}
\end{table}

As summarized in Table~\ref{tab:runtime}, the trajectory optimization stage dominates the total runtime, accounting for approximately $87\%$ of the per-epoch computation, while the grid-search MLE contributes about $13\%$ and the heap aggregation is negligible. Crucially, the total per-epoch runtime remains below $1$~s across all measured runs, including at the $99$th percentile ($834.98$~ms). This confirms that the complete computational pipeline supports the $1$~Hz update rate assumed in this work. We further note that the runtime can be reduced through more powerful hardware or additional parallelization, such as processing the per-UAV trajectory optimization in parallel.

We emphasize that the runtime reported here corresponds to the computational part of the closed-loop online pipeline. Detailed communication delay and network-level centralized aggregation latency depend on additional system-level assumptions, such as the communication protocol, bandwidth, packet loss, and network topology. These aspects are not explicitly modeled in the present work, which focuses on rigidity-based trajectory optimization for RSS-based localization. We therefore did not introduce an artificial communication-delay model, and we identify communication-aware implementation of the proposed framework as an important direction for future work.